%% file: ms.tex
\documentclass[11pt,a4paper]{article}
\pdfoutput=1
\usepackage{jcappub}

\usepackage{amsmath,amssymb,bm,float,Macro,mathrsfs,tabularx,color,graphics,hyperref}

\input{defs}

\title{Non-Gaussian Structure of B-mode Polarization after Delensing}

\author[a,b]{Toshiya Namikawa}
\author[c]{and Ryo Nagata}

\affiliation[a]{Department of Physics, Stanford University, Stanford, CA 94305, USA}
\affiliation[b]{Kavli Institute for Particle Astrophysics and Cosmology, SLAC National Accelerator
Laboratory, Menlo Park, CA 94025, USA}
\affiliation[c]{High Energy Accelerator Research Organization (KEK), Tsukuba, Ibaraki 305-0801, Japan}

\emailAdd{namikawa@slac.stanford.edu}
\emailAdd{rnagata@post.kek.jp}

\abstract{
The B-mode polarization of the cosmic microwave background on large scales 
has been considered as a probe of gravitational waves from the cosmic inflation. 
Ongoing and future experiments will, however, suffer from contamination due to the B-modes 
of non-primordial origins, one of which is the lensing induced B-mode polarization. 
Subtraction of the lensing B-modes, usually referred to as delensing, 
will be required for further improvement of detection sensitivity of the gravitational waves. 
In such experiments, knowledge of statistical properties of the B-modes after delensing is indispensable 
to likelihood analysis particularly because the lensing B-modes are known to be non-Gaussian. 
In this paper, we study non-Gaussian structure of the delensed B-modes on large scales, 
comparing it with that of the lensing B-modes. 
In particular, we investigate the power spectrum correlation matrix and the probability distribution 
function (PDF) of the power spectrum amplitude. 
Assuming an experiment in which the quadratic delensing is an almost optimal method, 
we find that delensing reduces correlations of the lensing B-mode power spectra between different multipoles,
and that the PDF of the power spectrum amplitude is well described as 
a normal distribution function with a variance larger than that in the case of a Gaussian field. 
These features are well captured by an analytic model based on the 4th order Edgeworth expansion. 
As a consequence of the non-Gaussianity, 
the constraint on the tensor-to-scalar ratio after delensing is degraded within approximately a few percent, 
which depends on the multipole range included in the analysis. 
}

\begin{document} 
\maketitle
\flushbottom

\input{sec1}

\input{sec2}

\input{sec3}

\input{sec4}

\input{sec5}

\input{sec6}

\acknowledgments 
TN is supported by JSPS fellowship for abroad (Grant No. 26-142). 
We acknowledge the use of 
{\tt Healpix} \citep{Gorski:2004by}, 
{\tt Lenspix} \citep{Challinor:2005jy} and 
{\tt CAMB} \citep{Lewis:1999bs}. 

\appendix
\input{appA}

\input{appB}

\bibliographystyle{mybst}
\bibliography{cite}

\end{document}

%% file: defs.tex
\def\grad{\phi}

\def\tX{\widetilde{X}}
\def\tY{\widetilde{Y}}
\def\rB{B^{\rm d}}

\def\CEE{C^{\rm EE}}

\def\Cgg{C^{\grad\grad}}
\def\Ngg{N^{\grad\grad}}
\def\hC{\widehat{C}}
\def\tC{\widetilde{C}}

\def\tCEE{\tC^{\rm EE}}
\def\hCBB{\hC^{\rm BB}}
\def\tCBB{\tC^{\rm BB}}
\def\rCBB{C^{\rm BB,d}}

\def\hr{\widehat{r}}
\def\estg{\widehat{\grad}}

\def\hE{\widehat{E}}
\def\hB{\widehat{B}}
\def\hA{\widehat{A}}
\def\hX{\widehat{X}}
\def\hY{\widehat{Y}}

\def\hCXX{\widehat{C}^{\rm XX}}

\def\hCYY{\widehat{C}^{\rm YY}}

\def\ol{\overline}

\def\bn{\bm{\nabla}}
\def\hatn{\hat{\bm{n}}}
\def\l{\ell}
\def\Cov{{\rm Cov}}

%% file: sec1.tex
\section{Introduction} \label{sec.1}

Measurement of the B-mode polarization of the cosmic microwave background (CMB) 
on angular scales larger than a few dozen arcminutes has been considered as 
the best avenue to probe the primordial gravitational waves. 
The joint analysis of BICEP, Keck and PLANCK reported the constraint on the tensor-to-scalar 
ratio as $r_{0.05}<0.12$ ($95$\% C.L.) \cite{Ade:2015tva}, 
while the much tighter constraint was recently obtained from PLANCK 2015 as 
$r_{0.002}<0.08$ ($95$\% C.L.) \cite{Ade:2015xua}, 
where the subscripts $0.05$ and $0.002$ are the pivot scale of the primordial power spectrum 
in unit of Mpc$^{-1}$. 
Still, there is no evidence for the presence of the primordial gravitational waves. 

On large scales, Galactic foreground emission, which is one of the most significant contaminations, 
is expected to dominate over the B-mode signals of the primordial gravitational waves 
(sometimes called as the primary B-modes). 
Many studies have been devoted for foreground rejection techniques 
\cite{Dunkley:2008am,Betoule:2009,Ichiki:2014}
and it would be possible to remove the foreground contamination sufficiently 
for the detection of the primordial gravitational waves of $r\sim 0.001$ 
\cite{Katayama:2011eh}. 
Even if the foreground contamination is successfully removed, 
there is still a significant contamination due to gravitational lensing from 
mass distribution between the last scattering surface and an observer. 
The gravitational lensing effect on the propagation of CMB photons 
disturbs the spatial pattern of the polarization map, 
which converts a small portion of E-modes into B-modes \cite{Zaldarriaga:1998ar}. 
The lensing B-modes on large scales has a spectrum like white noise 
and their amplitude is comparable to that of the primary B-modes of $r=0.01$ 
on the scales of the recombination bump ($\ell\sim 10$-$100$) \cite{Lewis:2006fu}. 
In addition to precise measurement of the lensing B-modes, 
subtraction of the lensing B-modes, 
usually referred to as {\it delensing}, will be required in ongoing and future CMB experiments. 
It is expected to improve detection sensitivity of the primary B-modes 
\cite{Seljak:2003pn,Smith:2010gu,Boyle:2014kba,Simard:2014aqa,Namikawa:2014lla}
and even the signals of other non-lensing sources 
such as cosmic strings (e.g. \cite{Teng:2011xc,Kamada:2014qta}), 
specific phenomena of modified gravity theories (e.g., \cite{Saltas:2014dha}), 
and self-ordering scalar fields \cite{Fenu:2009JCAP} (more generically, any cosmic defect network 
\cite{Figueroa:2013PRL}). 

For delensing, we need to know the lensing mass distribution, which is described by the lensing potential.
Measurement of the lensing potential became realized recently by multiple observations of 
CMB polarizations such as ACTPol \cite{vanEngelen:2014zlh}, PLANCK \cite{Ade:2015zua}, 
POLARBEAR \cite{PB1:2013a} and SPTpol \cite{Hanson:2013daa,Story:2014hni}. 
Near term and next generation CMB observations such as 
Advanced ACT \cite{Calabrese:2014gwa}, 
Simons Array \cite{SimonsArray}, 
SPT-3G \cite{Benson:2014},
CMB Stage-IV \cite{Abazajian:2013vfg}
will greatly improve sensitivity to the lensing potential. 
In addition, mass tracers at high redshifts such as the cosmic infrared background 
would be also able to be used for delensing \cite{Simard:2014aqa,Sherwin:2015baa}. 

To perform likelihood analysis of the B-modes, 
we should understand and characterize statistical properties of the lensing and delensed B-modes. 
The lensing B-modes are known to be non-Gaussian field, and 
their statistical properties and the impact on the cosmological parameter constraints 
have been discussed in several works \cite{Smith:2004up,Smith:2006nk,Li:2006pu,2012:Benoit}. 
These studies showed that off-diagonal correlations of the B-mode power spectrum 
lead to notable degradation in the cosmological parameter estimation. 
On the other hand, statistical properties of the delensed B-modes, 
the knowledge of which is indispensable to analyses in future B-mode experiments, has not been explored in detail. 

The main purpose of this paper is to explore how delensing changes the non-Gaussian properties 
of the lensing B-modes, in particular the off-diagonal elements of the power spectrum correlation matrix 
and the probability distribution function (PDF) of the power spectrum amplitude. 
We also discuss its impact on the estimation of the tensor-to-scalar ratio. 
To explore statistical properties of the delensed B-modes, 
we synthesize Monte Carlo samples of simulated CMB maps. 
Using the simulated maps, we estimate the lensing potentials and compute the delensed B-modes. 
The samples of the lensing and delensed B-modes are used for simulating 
the power spectrum covariance and the PDF of the power spectrum amplitude. 
To understand the simulation results, treating the delensed B-mode polarization 
as a weakly non-Gaussian field, we construct an analytic model which explains the simulated 
power spectrum covariance and also the simulated PDF of the power spectrum amplitude. 

This paper is organized as follows: 
In Sec.~\ref{sec.2}, we describe the procedures of our delensing analysis 
and generation of Monte Carlo samples. 
In Sec.~\ref{sec.3}, we show the simulation results, 
focusing on the power spectrum covariance and PDF of power spectrum amplitude of the B-mode polarization. 
In Sec.~\ref{sec.4}, we present analytic expressions for the statistics of the delensed B-modes 
and compare them with the simulation results. 
In Sec.~\ref{sec.5}, we show the impact of the non-Gaussianity on the estimation of 
the tensor-to-scalar ratio. 
Finally, Sec.~\ref{sec.6} is devoted to our conclusion and discussion.

Throughout this paper, we assume a flat $\Lambda$CDM model characterized by six parameters which are 
the baryon density ($\Omega\rom{b}h^2$), non-relativistic matter density ($\Omega\rom{m}h^2$), 
dark energy density ($\Omega_{\Lambda}$), scalar spectral index ($n\rom{s}$), 
scalar amplitude defined at $k=0.05$Mpc$^{-1}$ ($A\rom{s}$), and 
reionization optical depth ($\tau$). 
The cosmological parameters have the best-fit values of PLANCK 2013 results \cite{Ade:2013zuv}; 
$\Omega\rom{b}h^2=0.0220$, $\Omega\rom{m}h^2=0.1409$, $\Omega_{\Lambda}=0.6964$, 
$n\rom{s}=0.9675$, $A\rom{s}=2.215\times10^{-9}$, and $\tau=0.0949$.

%% file: sec2.tex
\section{Simulation Method} \label{sec.2}

In our analysis, to show non-Gaussian structure involved in the delensed B-mode power spectrum,
we first simulate lensed CMB polarization maps, and then perform quadratic lensing reconstruction and delensing. 
The number of our Monte Carlo samples is $10000$. 
In this section, we describe our method of simulating lensed polarization maps, 
quadratic lensing reconstruction and delensing. 
Note that the simulation method described below corresponds to our previous paper \cite{Namikawa:2014yca}.

\subsection{Map simulation}

We denote the polarization anisotropies at a position $\hatn$ on the last scattering surface 
as $[Q\pm\iu U](\hatn)$. The lensed polarization anisotropies in the direction $\hatn$, 
are then given by (e.g., \cite{Zaldarriaga:1998ar}):
\al{
	[Q\pm\iu U](\hatn) &= [Q\pm\iu U](\hatn + \bm{d}(\hatn)) \,, \label{Eq:remap}
}
where $\bm{d}$ is the deflection angle, and is given by the gradient of the lensing potential $\bn\grad$.
Here we ignore curl modes in our simulation (see e.g. \cite{Namikawa:2011cs}). 
Instead of being expressed as spin-$2$ quantities, the following E and B mode polarizations 
are usually analysed in harmonic space (e.g., \cite{Zaldarriaga:1998ar}):
\al{
	[E \pm \iu B ]_{\l m} = -\Int{}{\hatn}{_} {}_{\pm 2} Y_{\l m}^*(\hatn) [Q\pm \iu U](\hatn) 
	\,, 
}
where we denote the spin-$2$ spherical harmonics as ${}_{\pm 2}Y_{\l m}$. 
Similarly, with the spin-$0$ spherical harmonics, $Y_{\l m}$, 
the lensing potential is transformed into the harmonic space as
\al{
	\grad_{LM} = \Int{}{\hatn}{_} Y_{LM}^*(\hatn) \grad (\hatn) 
	\,.
}
To simulate the lensed CMB polarization map, we first compute the {\it unlensed} 
angular power spectra of the E-mode polarization ($\CEE_{\l}$)
and lensing potential ($\Cgg_L$) with {\tt CAMB} \cite{Lewis:1999bs}. 
In {\tt Lenspix} \cite{Challinor:2005jy}, according to these spectra, the harmonic coefficients, $E_{\l m}$ and 
$\grad_{LM}$, are generated as zero mean random Gaussian fields. 
Note that, in our lensing simulation, the primary B-mode polarization at the last scattering surface 
is not included since the lensed primary B-mode is much smaller than the lensing B-mode. 
Then, $E_{\l m}$ is transformed into Q and U maps while $\grad_{\l m}$ is 
transformed into $\grad(\hatn)$. 
Finally, the Q and U maps are remapped according to Eq.~\eqref{Eq:remap} using $\grad(\hatn)$. 
In our simulation, the Healpix pixelization parameter ({\tt nside}) is set to be $2048$. 

Expanding Eq.~\eqref{Eq:remap} up to the first order of the lensing potential, 
the B modes of the lensed polarization field are described as (e.g., \cite{Hu:2000ee}) 
\al{
	\tilde{B}_{\l m} &= - \iu\sum_{\l'm'}\sum_{LM} 
		\Wjm{\l}{\l'}{L}{m}{m'}{M} \mC{S}^{(-)}_{\l\l'L} E_{\l'm'}^* \grad_{LM}^* 
	\label{Eq:Lensed-B} \,, 
} 
where we ignore the primary B-mode. 
The quantity $\mC{S}^{(\pm)}_{\l\l'L}$ represents the mode coupling induced by the lensing: 
\al{ 
	\mC{S}^{(\pm)}_{\l\l'L} 
		&= p^{(\pm)}\sqrt{\frac{(2\l+1)(2\l'+1)(2L+1)}{16\pi}}
			[-\l(\l+1)+\l'(\l'+1)+L(L+1)]\Wjm{\l}{\l'}{L}{2}{-2}{0} 
	\,. \label{Eq:Spm}
} 
Here $p^{(+)}$ ($p^{(-)}$) is unity if $\l+\l'+L$ is an even (odd) integer and zero otherwise. 
Eq.~\eqref{Eq:Lensed-B} is known to be a good analytic approximation for the lensing B-modes on large scales. 
From Eq.~\eqref{Eq:Lensed-B}, the lensing B-modes in the absence of the primary B-modes are simply 
expressed in terms of a convolution between the E-mode polarization and lensing potential as
\al{
	\widetilde{B}_{\l m} = \mS{B}_{\l m}[E,\grad]
	\,, \label{Eq:Lensing-E-to-B}
}
where we simplify Eq.~\eqref{Eq:Lensed-B} by defining a convolution operator for two multipole moments:
\al{
	\mS{B}_{\l m}[\alpha,\beta] \equiv -\iu\sum_{\l'm'}\sum_{LM} 
		\Wjm{\l}{\l'}{L}{m}{m'}{M}\mC{S}^{(-)}_{\l\l'L} \alpha_{\l'm'}^*\beta_{LM}^*
	\,. 
}

\subsection{Quadratic lensing reconstruction} 

From the lensed polarization map, we reconstruct the lensing potential $\grad$ with the following method. 
Lensing induces the off-diagonal elements of the covariance matrix 
($\l\not=\l'$ or $m\not=m'$) between two lensed polarization anisotropies ($X,Y=E,B$) as 
\al{
	\ave{\tX_{\l m}\tY_{\l'm'}}\rom{CMB} 
		&= \sum_{L,M}\Wjm{\l}{\l'}{L}{m}{m'}{M}f^{\rm XY}_{\l\l'L}\grad^*_{LM} + \mC{O}(\grad^2)
	\,, \label{Eq:weight} 
}
where the operation $\ave{\cdots}\rom{CMB}$ denotes the ensemble average over the primary CMB anisotropies. 
The weight function $f_{\l\l'L}^{\rm XY}$ in Eq.~\eqref{Eq:weight} is defined as \cite{Okamoto:2003zw} 
\al{
	f^{\rm EE}_{\l\l'L} &= \mC{S}^{(+)}_{\l\l'L}\CEE_{\l'} + \mC{S}^{(+)}_{\l'\l L}\CEE_{\l} 
	\,, \\
	f^{\rm EB}_{\l\l'L} &= \iu \mC{S}^{(-)}_{\l'\l L}\CEE_{\l}
	\,. \label{Eq:f}
}
Eq.~\eqref{Eq:weight} motivates the following form for a quadratic lensing estimator
(see e.g., \cite{Okamoto:2003zw}):
\al{
	[\estg^{\rm XY}_{LM}]^* 
		= A^{\rm XY}_L\sum_{\l\l'}\sum_{mm'}\Wjm{\l}{\l'}{L}{m}{m'}{M}
		\frac{1}{\Delta^{\rm XY}}[f^{\rm XY}_{\l\l'L}]^*\frac{\hX_{\l m}}{\hCXX_{\l}}\frac{\hY_{\l'm'}}{\hCYY_{\l'}}
	\,, \label{Eq:estg-XY}
}
where $\hX$ and $\hY$ are observed polarization anisotropies. 
The quantities, $\hCXX_{\l}$ and $\hCYY_{\l}$, are their angular power spectra. 
$\Delta^{\rm XY}$ is $2$ for ${\rm XY}={\rm EE}$ and $1$ for ${\rm XY}={\rm EB}$. 
The quantity $A^{\rm XY}_L$ is given by 
\al{
	A^{\rm XY}_L &= \left\{\frac{1}{2L+1}\sum_{\l\l'} 
		\frac{[f^{\rm XY}_{\l\l'L}]^*f^{\rm XY}_{\l\l'L}}{\Delta^{\rm XY}\hCXX_{\l}\hCYY_{\l'}}\right\}^{-1}
	\,. \label{Eq:Rec:N0}
}
Note that, to mitigate biases coming from the higher order terms of $\grad$, 
we use the lensed power spectrum ($\tCEE_{\l}$) in Eq.~\eqref{Eq:f} \cite{Hanson:2010rp,Lewis:2011fk} 
instead of the unlensed power spectrum. 
The optimal estimator of the lensing potential is then obtained as a linear combination of 
the EE and EB quadratic estimators: 
\al{
	\estg_{LM} = A^{\grad}_L\left(\frac{1}{A_L^{\rm EE}}\estg_{LM}^{\rm EE} 
		+ \frac{1}{A_L^{\rm EB}}\estg_{LM}^{\rm EB}\right)
	\,, \label{Eq:estg}
}
with
\al{
	A_L^{\grad} \equiv \frac{1}{(A_L^{\rm EE})^{-1}+(A_L^{\rm EB})^{-1}}
	\,. \label{Eq:AL_GRAD}
}

In lensing reconstruction, we add a random white noise with a Gaussian beam factor  
whose power spectrum is described as
\al{
	N^{\rm P}_{\l} &\equiv \left(\frac{\Delta\rom{P}}{T\rom{CMB}}\right)^2
		\exp\left[\frac{\l(\l+1)\theta^2}{8\ln 2}\right]
	\,. \label{noise}
}
Here $T\rom{CMB}=2.7$K is the mean temperature of CMB, 
$\theta$ is a beam size, and $\Delta\rom{P}$ is a  noise level of polarization measurement. 
The multipole $\l$ is an integer between $2$ and $2000$. 
The fiducial values for the experiment assumed in our analysis are $\Delta\rom{P}=6\mu$K-arcmin and 
$\theta=4$ arcmin. In such a situation, quadratic lensing reconstruction described here is 
a nearly optimal method \cite{Hirata:2003ka}. 
We estimate the lensing potential, using the E and B-modes up to $\l=2000$. 
Note that these values are similar to those of the Simons Array \cite{SimonsArray}
which is expected to cover a wide range of the entire sky. 
Advanced ACT also has a similar instrumental specification \cite{Calabrese:2014gwa}. 

\subsection{Quadratic delensing} 

Once we obtain the lensing potential using the quadratic estimator, 
we can estimate the lensing B-modes in a similar manner of Eq.~\eqref{Eq:Lensing-E-to-B}. 
On large scales, the lensing B-mode is well estimated by \cite{Smith:2010gu}: 
\al{
	\mS{B}_{\l m}[\hE^{\rm w},\estg^{\rm w}]
	\,, \label{Eq:Quad-LensB}
}
where, again, the operator $\mS{B}_{\l m}$ convolves two fields in the same way that 
it forms the lensing B-modes from the E-modes and lensing potential. 
We denote $\hE^{\rm w}_{\l m}$ and $\estg^{\rm w}_{\l m}$ as the Wiener filtered multipoles, i.e., 
the measured E-mode polarization and lensing potential multiplied by the corresponding Wiener filters ($W^{\rm E}$ 
and $W^{\grad}$) defined as \cite{Seljak:2003pn,Smith:2010gu}
\al{
	W^{\rm E}_{\l} &= \frac{\CEE_{\l}}{\CEE_{\l}+N_{\l}^{\rm P}} 
	\,, \\
	W^{\grad}_{\l} &= \frac{\Cgg_{\l}}{\Cgg_{\l}+A_{\l}^{\grad}} 
	\,. \label{Eq:WF-grad}
}
The residual field of the B-mode polarization after delensing ($\rB_{\l m}$) is then evaluated as \cite{Smith:2010gu}
\al{
	\rB_{\l m} = \hB_{\l m} - \mS{B}_{\l m}[\hE^{\rm w},\estg^{\rm w}] 
	\,. \label{Eq:Quad-Delens}
}
To avoid delensing bias (see e.g. appendix A of Ref.~\cite{Namikawa:2014yca}), 
only the multipoles between $\l=301$ and $2000$ are taken into account in the calculation of lensing reconstruction. 
Then, we evaluate the delensed B-mode polarization only between $\l=2$ and $300$. 
Note that, as shown in our previous paper \cite{Namikawa:2014yca}, the filtering of large scale 
multipoles in lensing reconstruction is required also for an unbiased estimate of 
the lensing potential in presence of $1/f$ noises.

%% file: sec3.tex
\section{Non-Gaussian Signature of Delensed B-mode Polarization} \label{sec.3}

\subsection{Power spectrum covariance} 

\begin{figure} 
\bc
\includegraphics[width=8.8cm,clip]{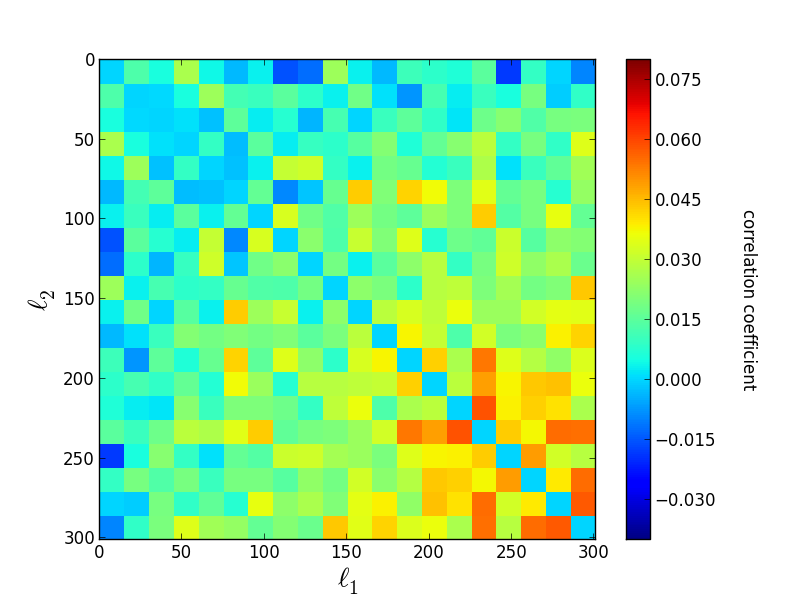}
\hspace{-6.2em}
\includegraphics[width=8.8cm,clip]{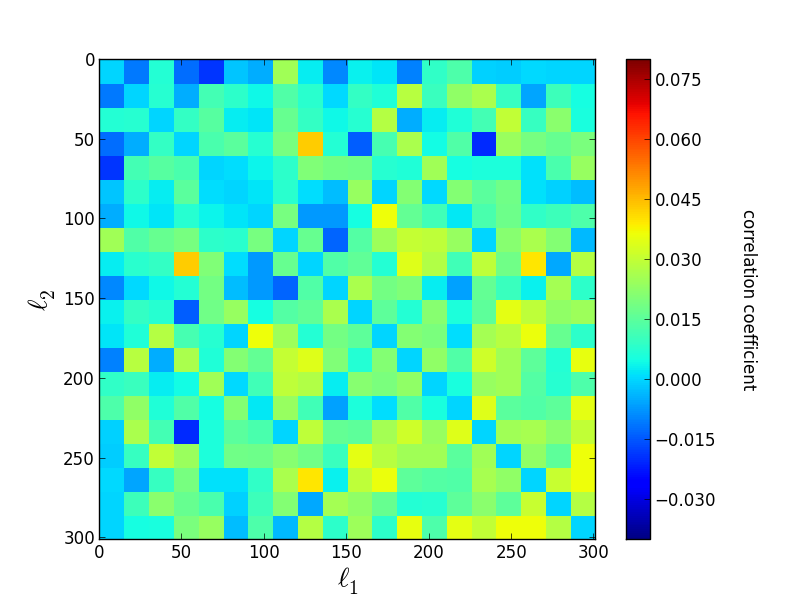}
\caption{
Correlation coefficients of the angular power spectrum of the lensing (Left) and 
delensed (Right) B-modes between different multipoles defined in Eq.~\eqref{Eq:Rij}. 
The correlation coefficients are computed from $10000$ realizations of the lensed and delensed QU maps. 
We apply flat binning. The bin width is $\Delta\l=15$ and 
the multipoles between $\l=1$ and $300$ are divided into $20$ bins. 
Note that the diagonal elements are removed for an illustrative purpose. 
}
\label{Fig:Cov}
\ec
\end{figure}

Significance of non-Gaussian properties of the B-mode polarization can be seen in 
the power spectrum covariance. 
In the case of a Gaussian field, the power spectra at different multipoles are 
uncorrelated, and the covariance matrix has only diagonal elements. 
On the other hand, a non-Gaussian field such as the lensing B-modes has off-diagonal elements of 
notable magnitude in the covariance matrix as shown in the previous studies 
\cite{Smith:2004up,Smith:2006nk,Li:2006pu,2012:Benoit}.
Here, we present the simulated covariance matrix of the delensed B-mode power spectrum and compare it with 
that of the lensing B-mode power spectrum. Note that, to see non-Gaussian structure of 
the delensed B-modes, we do not include the tensor perturbations at first. 
The case including the tensor perturbations is discussed in Sec.~\ref{sec.5}. 

To compute the power spectrum covariance, the multipole is binned into $20$ bins. 
The binned angular power spectrum at $i$th multipole bin is defined as
\al{
	\hC_i = \frac{1}{N_i}\sum_{\l\in [\l_{i-1},\l_i-1]} \hC_{\l}
	\,,
}
where $N_i$ is the number of multipoles in $i$th bin, and we choose $\l_i=15i+1$. 
For the first bin ($i=1$), we remove $\l=1$. 
The angular power spectrum is computed with the usual method:
\al{
	\hC_{\l} = \frac{1}{2\l+1}\sum_{m=-\l}^{\l}|B_{\l m}|^2
	\,.
}
The power spectrum covariance between each multipole bin $i$ and $j$ is then evaluated as
\al{
	\Cov_{ij} = \ave{\hC_{i}\hC_{j}} - \ave{\hC_{i}}\ave{\hC_{j}}
	\,. 
}
$\ave{\cdots}$ is the sample mean of the operand. 

In Fig.~\ref{Fig:Cov}, we show the correlation coefficients of 
the lensing (Left) and delensed (Right) B-mode power spectrum defined as \cite{2012:Benoit}
\al{
	R_{ij} = \frac{\Cov_{ij}}{\sqrt{\Cov_{ii}\Cov_{jj}}}
	\,. \label{Eq:Rij}
}
The values of the off-diagonal elements increase at higher multipoles. The typical values of 
the correlation coefficients of the lensing B-modes are $\sim 6-8$\% at $\l\sim 200-300$, 
which is consistent with the previous results obtained in Ref.~\cite{2012:Benoit}
\footnote{
Note that Ref.~\cite{2012:Benoit} showed the correlation coefficients with $2$ times wider bins 
in which the off-diagonal elements are apparently larger than those in Fig.~\ref{Fig:Cov}. 
}. 
On the other hand, we can confirm that the off-diagonal elements of the correlation matrix of the delensed B-modes 
are definitely smaller than those of the lensing B-modes. These results imply that the delensing operation 
reduced the off-diagonal covariance more efficiently than the diagonal covariance. 

\subsection{Probability distribution function of power spectrum amplitude}

\begin{figure} 
\bc
\includegraphics[width=7.8cm,clip]{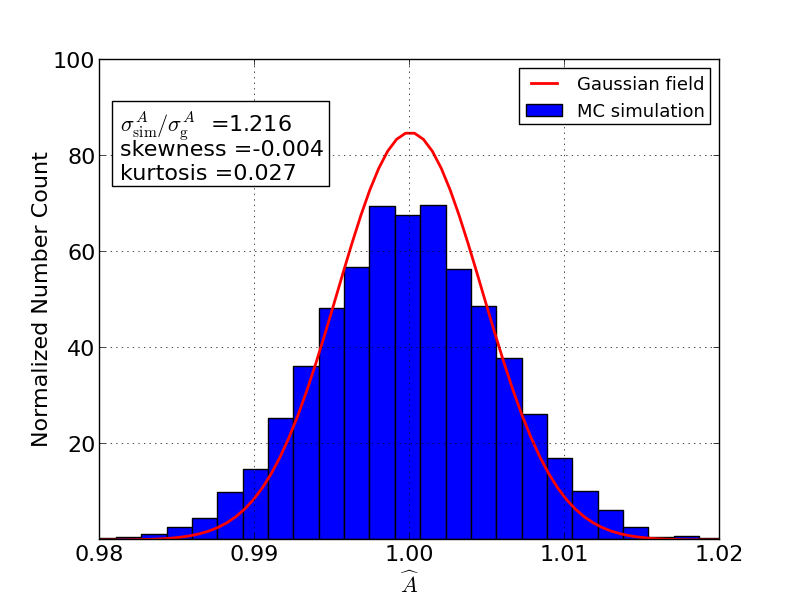}
\hspace{-2em}
\includegraphics[width=7.8cm,clip]{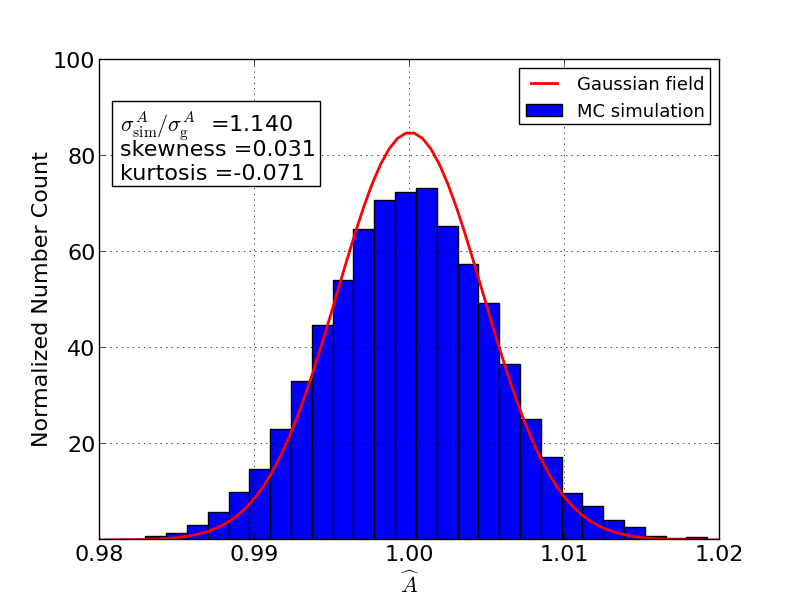}
\caption{
Histogram of the estimated power spectrum amplitude of the lensing (Left) and delensed (Right) B-modes. 
The number of the samples is $10000$. The solid line shows the PDF obtained on the assumption that 
the lensing/delensed B-modes obey Gaussian statistics. We show the ratio of the standard deviation 
computed from the simulation ($\sigma\rom{sim}^A$) to that in the case of a Gaussian CMB field 
($\sigma\rom{g}^A$). The skewness and kurtosis of the histogram are also shown. 
The fiducial power spectrum is evaluated as the sample mean of the simulated power spectra. 
}
\label{Fig:Al}
\ec
\end{figure}

To discuss the non-Gaussian covariance more quantitatively, next we consider the estimation of 
the power spectrum amplitude. In the case of a Gaussian field, the optimal estimator of 
the power spectrum amplitude is given by \cite{Hamimeche:2008ai}
\al{
	\hA = \frac{\sum_{\l}(2\l+1)\hA_{\l}}{\sum_{\l}(2\l+1)}
	\,, \label{Eq:Est-Amp}
}
where $\hA_{\l}$ is the amplitude parameter at each multipole:
\al{
	\hA_{\l} = \frac{\hC_{\l}}{C_{\l}} 
	\,. 
}
The angular power spectrum in the denominator ($C_{\l}$) is the fiducial power spectrum defined as 
the ensemble average of $\hC_{\l}$. 
The parameter $\hA_{\l}$ is a measured amplitude relative to $C_{\l}$ at each multipole. 
In the case of a Gaussian CMB field, the expected variance of $\hA$ is given by 
\al{
	(\sigma^A\rom{g})^2 = \frac{2}{\sum_{\l}(2\l+1)}
	\,. \label{Eq:varA-g}
}
Deviation of the simulated variance from $(\sigma^A\rom{g})^2$ indicates a non-Gaussian signature. 

The histograms of the estimator $\hA$ \eqref{Eq:Est-Amp} of the lensing and delensed B-modes are shown 
in Fig.~\ref{Fig:Al}. They are synthesized from $10000$ realizations. 
In the estimation of $\hA$, we use the B-modes between $\l=2$ and $300$.
We evaluate the fiducial power spectrum as the sample mean of the simulated $10000$ power spectra. 
Consequently, the sample mean of $\hA$ is unbiased. The standard deviation of $\hA$ ($\sigma^A\rom{sim}$) 
divided by that in the case of a Gaussian field ($\sigma^A\rom{g}$) is shown in the figure. 

Before delensing, the standard deviation of the estimator is $\sim 22$\% larger than that in the case of 
a Gaussian CMB field. On the other hand, after delensing, the discrepancy is decreased to $\sim 15$\%, 
which is considered to be an outcome of the decrease of the off-diagonal correlations. 
Note that, even in the case that the lensing/delensed B-modes are assumed to be Gaussian, 
the skewness and kurtosis of the amplitude estimator have non-zero values because the amplitude estimator 
is quadratic in the CMB field. In our case, the skewness and kurtosis of the estimator are 
$0.0057$ and $1.3\times 10^{-7}$, respectively. The skewness and kurtosis are also shown, and they 
deviate from those in the case of a Gaussian CMB field within only $\leq 0.03$ for the skewness and 
$\leq 0.07$ for the kurtosis, respectively.

%% file: sec4.tex
\section{Modeling Statistical Properties of Delensed B-mode} \label{sec.4}

To understand the statistical properties of the delensed B-modes shown in the previous section, 
we construct an analytic model of the PDF of the lensing and delensed B-modes which is consistent 
with the simulation results. 
The simulation results suggest that the deviation from a Gaussian field is appeared 
primarily as the small increase of the variance $(\sigma^A)^2$. 
Motivated by this fact, we try to express the PDF of the lensing and delensed B-modes 
as a perturbed Gaussian distribution.
The Edgeworth expansion, which assumes a weakly non-Gaussian PDF, is used for such purpose.
We derive a PDF for the lensing and delensed B-modes 
which includes corrections up to 4th order of the Edgeworth expansion.

\subsection{Probability distribution function for weakly non-Gaussian fields} \label{Sec:PDF-general}

We denote $n_L(=\sum_{\l}(2\l+1))$ independent Gaussian random variables as $a_{\l m}$. 
The joint PDF of $\bm{a}=\{a_{\l m}\}$ is given by
\al{
	P\rom{g}(\bm{a}) = \frac{1}{(2\pi)^{n_L/2}|\bR{C}|^{1/2}}
		\exp\left[-\frac{1}{2}\bm{a}^t\bR{C}^{-1}\bm{a}\right]
	\,. 
}
Here, we denote the covariance of $a_{\l m}$s as $\bR{C}=\ave{\bm{a}\bm{a}^t}$. 
The PDF including the corrections from the Edgeworth expansion at 4th order is given by 
\al{
	P(\bm{a}) = [1+k(\bm{a})] P_{\rm g}(\bm{a}) 
	\,, 
}
where the 4th-order correction term $k(\bm{a})$ is described as (see e.g. \cite{2010:Regan})
\al{
	k(\bm{a}) 
		&= \frac{1}{24 P\rom{g}(\bm{a})}\sum_{\l_i m_i} T^{\l_1\l_2\l_3\l_4}_{m_1m_2m_3m_4}
			\PD{}{a_{\l_1 m_1}}\PD{}{a_{\l_2 m_2}}\PD{}{a_{\l_3 m_3}}\PD{}{a_{\l_4 m_4}}
			P\rom{g}(\bm{a})
	\label{Eq:deriv-k4} \\
		&= \frac{1}{24}\sum_{\l_i m_i} T^{\l_1\l_2\l_3\l_4}_{m_1m_2m_3m_4}
		\bigg(
			\ol{a}_{\l_1 m_1}\ol{a}_{\l_2 m_2}\ol{a}_{\l_3 m_3}\ol{a}_{\l_4 m_4}
	\notag \\ 
		&\qquad
			- 6 \bR{C}^{-1}_{\l_1 m_1, \l_2 m_2}\ol{a}_{\l_3 m_3}\ol{a}_{\l_4 m_4}
			+ 3 \bR{C}^{-1}_{\l_1 m_1, \l_2 m_2}\bR{C}^{-1}_{\l_3 m_3, \l_4 m_4}
		\bigg)
	\label{Eq:sum-k4} 
	\,. 
}
The quantity $\ol{a}_{\l m}=[\bR{C}^{-1}\bm{a}]_{\l m}$ is the inverse-variance filtered multipole. 
We also denote the connected part of the four-point correlation as 
\al{
	T^{\l_1\l_2\l_3\l_4}_{m_1m_2m_3m_4}
		= \ave{a_{\l_1 m_1}a_{\l_2 m_2}a_{\l_3 m_3}a_{\l_4 m_4}}_c 
	\,. 
}
We assume that the covariance matrix of $a_{\l m}$ is diagonal:
\al{
	&\bR{C}_{\l m,\l'm'} = \delta_{\l\l'}\delta_{m,-m'}(-1)^m C_{\l} 
	\,. 
}
This simplifies the PDF as 
\al{
	P(\bm{a}) &= \frac{1+k(\bm{a})}{(2\pi)^{n_L/2}} \prod_{\l} C^{-\frac{2\l+1}{2}}_{\l}
		\exp\left[-\sum_{m=-\l}^{\l}\frac{|a_{\l m}|^2}{2C_{\l}}\right]
	\,. \label{Eq:PDF}
}
Note that the PDF described above is used to derive the optimal trispectrum estimator of 
the lensing potential power spectrum \cite{Namikawa:2012pe,Namikawa:2013xka}. 

In the above PDF, we ignore the third order term in the Edgeworth expansion which is expressed 
in terms of the B-mode bispectrum because the bispectrum of the B-mode polarization has 
the odd-parity symmetry and is not generated in the standard cosmology.

Let us discuss the statistical property of the power spectrum amplitude, $\hA$. 
Using the PDF $P(\bm{a})$ given by Eq.~\eqref{Eq:PDF}, we obtain the mean of the estimator as follows:
\al{
	\ave{\hA} = \ave{(\hA + \hA k(\bm{a}))}\rom{g} = 1
	\,.
}
Here, the operation $\ave{\cdots}$ means averaging based on the whole PDF $P(\bm{a})$, 
while the operation $\ave{\cdots}\rom{g}$ means that on the Gaussian PDF $P\rom{g}(\bm{a})$. 
The derivation is shown in appendix~\ref{sec:amp-mean}. 
On the other hand, from the PDF described in Eq.~\eqref{Eq:PDF}, 
the variance of the amplitude estimator is given by (see appendix~\ref{sec:amp-var})
\al{
	(\sigma^A)^2 \equiv \ave{\hA^2} - 1
		&= \frac{(\sigma^A\rom{g})^2}{n_L}\sum_{\l\l'}\frac{(2\l+1)(2\l'+1)}{2}
		\frac{{\rm Cov}_{\l\l'}}{C_{\l}C_{\l'}} 
	\,, \label{Eq:sigmaA}
}
where the power spectrum covariance, Cov$_{\l\l'}$, is expressed as
\al{
	{\rm Cov}_{\l\l'} &= \frac{2C_{\l}C_{\l'}}{2\l+1}\delta_{\l\l'}
		+ \frac{1}{(2\l+1)(2\l'+1)}\sum_{mm'} (-1)^{m+m'}T^{\l\l\l'\l'}_{m,-m,m',-m'}
	\,. \label{Eq:Cov-Full}
}
The PDF of the power spectrum amplitude $\hA$ is given as the chi square distribution with 
a correction term from the kurtosis (see appendix \ref{sec:amp-PDF} for derivation): 
\al{
	P(\hA) = \frac{(n_L/2)^{n_L/2}}{\Gamma(n_L/2)} \hA^{\frac{n_L}{2}-1} \E^{-\frac{n_L}{2}\hA} 
		\left\{1+n_L\frac{(\sigma^A)^2-(\sigma^A\rom{g})^2}{4(\sigma^A\rom{g})^2}
		\left[\frac{n_L}{n_L+2}\hA^2-2\hA+1\right]\right\}
	\,. \label{Eq:PDF-analytic}
}
Once we obtain the power spectrum covariance, we can actually evaluate the variance 
of the power spectrum amplitude and the PDF. 

\subsection{Expression for B-mode power spectrum covariance} 

Next, we discuss mathematical expression for the B-mode power spectrum covariance.
At first, we summarize the derivation of the analytic expression for the power spectrum covariance of 
the lensing B-modes developed by Refs.~\cite{Smith:2004up,Smith:2006nk,Li:2006pu,2012:Benoit}. 
The lensing B-mode power spectrum on large scales ($\l\lsim 300$) 
is well described by a convolution of the E-mode polarization
and lensing potential power spectra (see e.g., \cite{Smith:2010gu}):
\al{
	\tCBB_{\l} = \frac{1}{2\l+1}\sum_{\l'L} (\mC{S}^{(-)}_{\l\l'L})^2 \CEE_{\l'}\Cgg_L
		 \equiv \Xi_{\l}[\CEE,\Cgg]
	\,,
}
where $S^{(-)}_{\l\l'L}$ is given in Eq.~\eqref{Eq:Spm}, and we define a convolution operator:
\al{
	\Xi_{\l}[A,B] = \frac{1}{2\l+1}\sum_{\l_1\l_2} (\mC{S}^{(-)}_{\l\l_1\l_2})^2 A_{\l_1}B_{\l_2}
	\,. 
}
The fluctuations of the E-mode and lensing potential power spectra contribute to the fluctuation of 
the lensing B-mode power spectrum as follows \cite{2012:Benoit}: 
\al{
	\delta \tCBB_{\l} = \sum_{\l'}\PD{\tCBB_{\l}}{\CEE_{\l'}}\delta \CEE_{\l'}
		+ \sum_{\l'}\PD{\tCBB_{\l}}{\Cgg_{\l'}} \delta \Cgg_{\l'}
	\,. 
}
The connected part of the covariance due to this contribution $\ave{\tCBB_{\l}\tCBB_{\l'}}\rom{c}$ 
is approximately given by
\al{
	\ave{\tCBB_{\l}\tCBB_{\l'}}\rom{c} =  
		\sum_L \PD{\tCBB_{\l}}{\CEE_L}\frac{2(\CEE_L)^2}{2L+1}\PD{\tCBB_{\l'}}{\CEE_L}
		+ \sum_L \PD{\tCBB_{\l}}{\Cgg_L}\frac{2(\Cgg_L)^2}{2L+1}\PD{\tCBB_{\l'}}{\Cgg_L}
	\,. \label{Eq:cov-lensed}
}
Here we use 
\al{
	\ave{\delta C^X_{\l}\delta C^X_{\l'}} = \frac{2C^X_{\l}C^X_{\l'}}{2\l+1}\delta_{\l\l'}
	\,,
}
where $X$ is $EE$ or $\phi\phi$. Note that the power spectrum covariance also contains 
a fully connected term which is not expressed like the r.h.s. of Eq.~\eqref{Eq:cov-lensed}, 
but it has only negligible contribution \cite{Li:2006pu}. 
The approximate expression for the power spectrum covariance is then described as \cite{2012:Benoit} 
\al{
	{\rm Cov}_{\l\l'}^{\rm BB} 
	&\equiv \ave{\tCBB_{\l}\tCBB_{\l'}} - \ave{\tCBB_{\l}}\ave{\tCBB_{\l'}} 
	\notag \\
	&\simeq \frac{2(\tCBB_{\l})^2}{2\l+1}\delta_{\l\l'}
		+ \sum_L\left[\PD{\tCBB_{\l}}{\CEE_L}\frac{2(\CEE_L)^2}{2L+1}\PD{\tCBB_{\l'}}{\CEE_L}
		+ \PD{\tCBB_{\l}}{\Cgg_L}\frac{2(\Cgg_L)^2}{2L+1}\PD{\tCBB_{\l'}}{\Cgg_L}\right]
	\,. \label{Eq:CovBB-l}
}

We now apply the above discussion to derive the power spectrum covariance for the delensed B-modes. 
Assuming that the E-modes are nearly cosmic-variance limited, and using the similar procedure 
described above, we obtain the following expression as 
\al{
	{\rm Cov}^{\rm BB,d}_{\l\l'} = \frac{2(\rCBB_{\l})^2}{2\l+1}\delta_{\l\l'}
		+ \sum_L \left[\PD{\rCBB_{\l}}{\CEE_L}\frac{2(\CEE_L)^2}{2L+1}\PD{\rCBB_{\l'}}{\CEE_L}
		+ \PD{\rCBB_{\l}}{\Cgg_L}\frac{2(\Cgg_L)^2}{2L+1}\PD{\rCBB_{\l'}}{\Cgg_L}\right]
	\,. \label{Eq:CovBB-d}
}
The actual derivation requires cumbersome calculation and is shown in appendix \ref{sec:covBB}. 

\subsection{Comparison with simulation results}

Now we compare the model calculations with the results of the Monte Carlo simulation. 

\begin{figure} 
\bc
\includegraphics[width=7.8cm,clip]{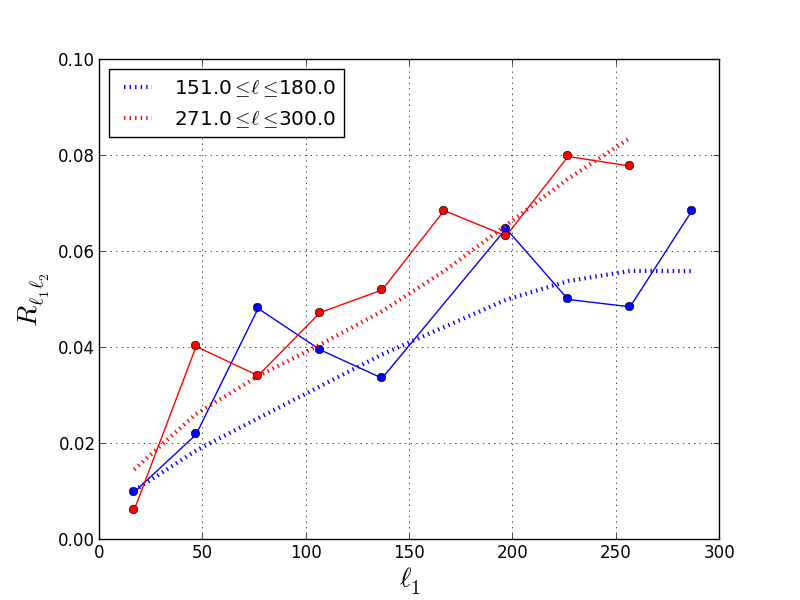}
\hspace{-2em}
\includegraphics[width=7.8cm,clip]{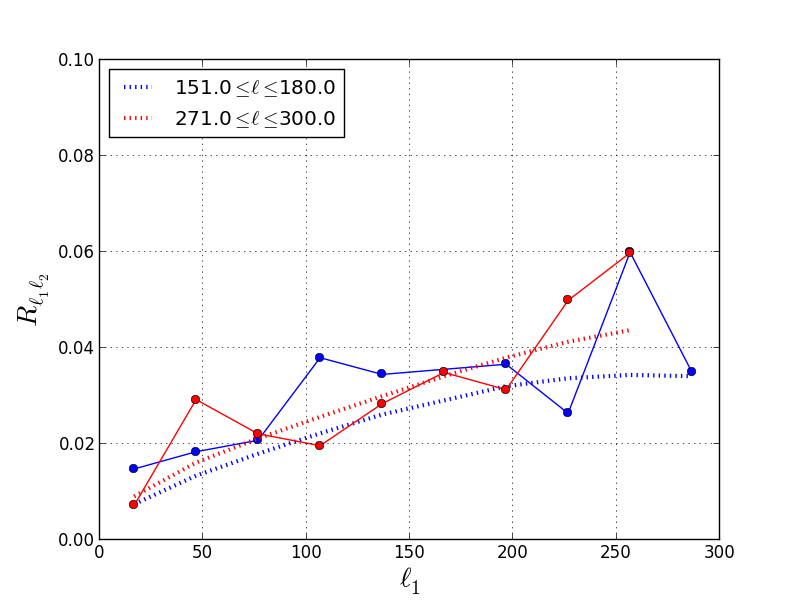}
\caption{
Correlation coefficients of the lensing (Left) and delensed (Right) B-mode power spectrum. 
The correlation coefficients obtained from the Monte Carlo simulation (solid) are compared 
with those of the analytic model (dashed). 
For an illustrative purpose, the multipoles up to $300$ are binned into $12$ bins. 
}
\label{Fig:R}
\ec
\end{figure}

\begin{figure} 
\bc
\includegraphics[width=7.8cm,clip]{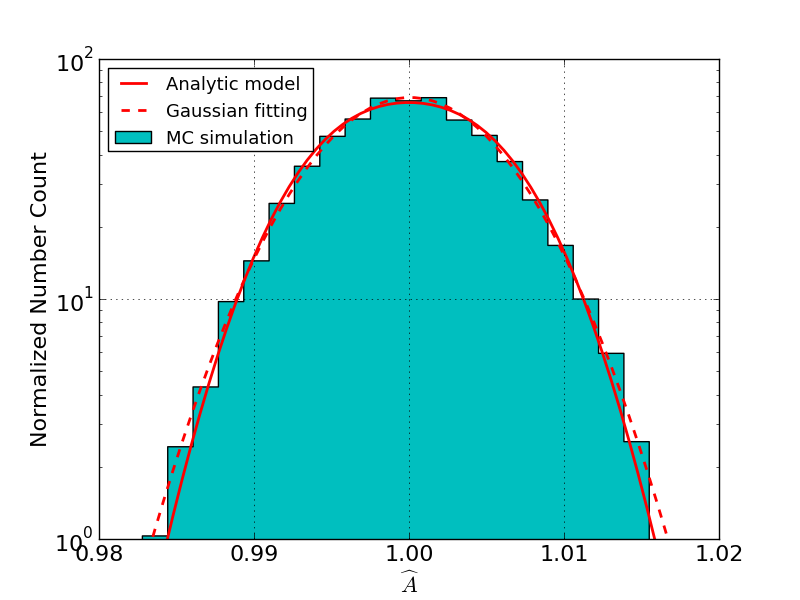}
\hspace{-2em}
\includegraphics[width=7.8cm,clip]{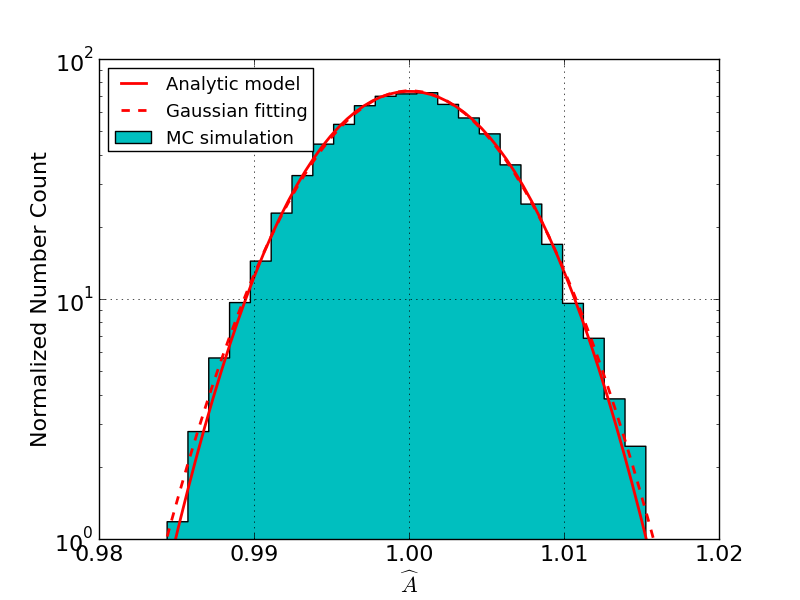}
\caption{
Theoretical PDF of the lensing (Left) and delensed (Right) B-mode power spectrum amplitude 
compared with the histogram shown in Fig.~\ref{Fig:Al}. 
We show two theoretical models; the analytic model of the PDF given by Eq.~\eqref{Eq:PDF-analytic} 
(solid), and the empirical PDF obtained by fitting a normal distribution with the histogram (dashed). 
Note that, the y-axis is in logarithmic scale for the clarification of the effect on the variance. 
}
\label{Fig:Al-comp}
\ec
\end{figure}

The correlation coefficients of the lensing and delensed B-mode power spectra are shown 
in Fig.~\ref{Fig:R}. 
The simulated correlation coefficients are compared with those of the analytic models described 
in the previous subsection. 
The analytic models capture the behavior of the simulated correlation coefficients. 

The analytic PDFs of the power spectrum amplitude described in Sec.~\ref{Sec:PDF-general} are compared 
with the histograms obtained from the Monte Carlo simulation in Fig.~\ref{Fig:Al-comp}. 
Our analytic model of the delensed B-modes well describes the statistics of the simulated samples. 
The simulated standard deviation of the amplitude parameter is $\sigma^A=0.005357$. 
By use of the analytic formula, the standard deviation is estimated as  
$\sigma^A=0.005264$ which is in agreement with the simulation result within $1.8$\%.
We also show the normal distributions fitted with the histograms. 
Contrary to the PDFs in the case of a Gaussian field (Fig.~\ref{Fig:Al}), 
the normal distributions with the fitted variances well capture the behaviors of the simulated histograms, 
which also validates the use of the 4th-order Edgeworth expansion.

%% file: sec5.tex
\section{Impact on Estimation of Tensor-to-Scalar Ratio} \label{sec.5}

\begin{figure} 
\bc
\includegraphics[width=7.8cm,clip]{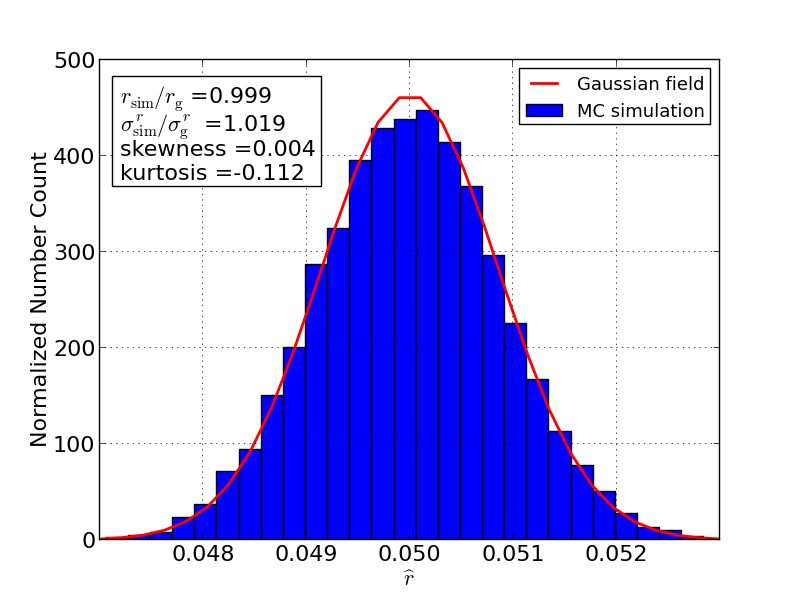}
\hspace{-2em}
\includegraphics[width=7.8cm,clip]{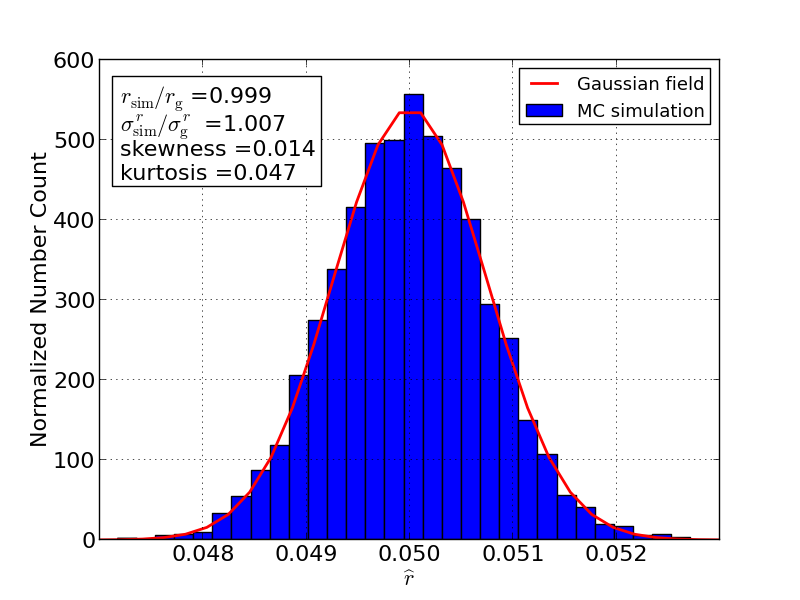}
\caption{
Histogram of the tensor-to-scalar ratio estimated from the tensor plus lensing (Left) and tensor plus 
delensed (Right) B-modes. The number of the samples is $10000$. The input value of 
the tensor-to-scalar ratio is $r=0.05$. The solid line shows the PDF in the case of a Gaussian B-mode. 
The fiducial power spectrum is evaluated as the sample mean of the simulated power spectra. 
}
\label{Fig:r1}
\ec
\end{figure}

\begin{figure} 
\bc
\includegraphics[width=10cm,clip]{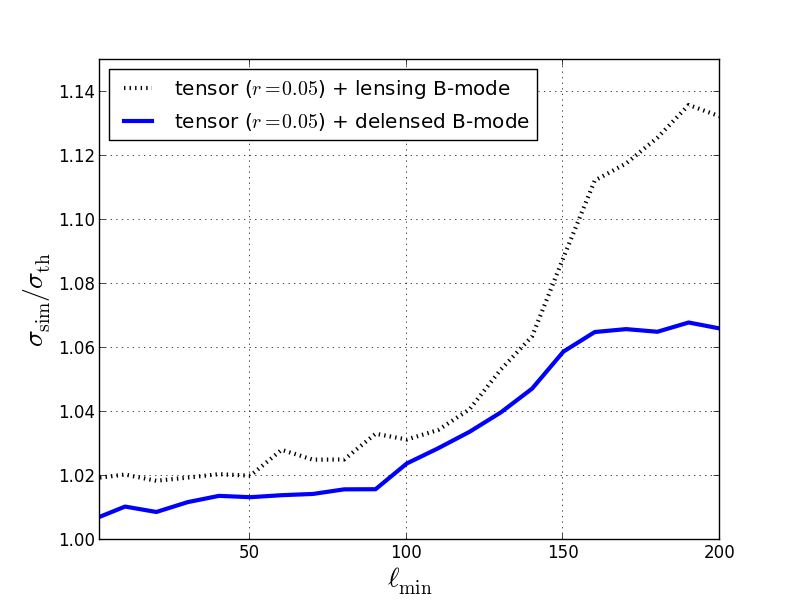}
\caption{
Standard deviations of the tensor-to-scalar ratio estimator divided by that in the case of 
a Gaussian B-mode ($\sigma^r_{\rm sim}/\sigma^r\rom{g}$) as a function of the minimum multipole 
included in the estimation ($\l\rom{min}$) for the tensor plus lensing (dashed black) and 
tensor plus delensed (solid blue) B-modes. The input value of the tensor-to-scalar ratio is $0.05$. 
The maximum multipole in the estimation is fixed to $300$. 
}
\label{Fig:r2}
\ec
\end{figure}

Here we demonstrate the effect of non-Gaussianity from the lensing/delensed B-modes to the estimation of 
the tensor-to-scalar ratio $r$. 
At a single multipole $\l$, the tensor-to-scalar ratio which maximizes the likelihood function 
is obtained by, e.g., differentiating Eq.~(3) of \cite{Katayama:2011eh} in terms of $r$:
\al{
	\hr_{\l} = \frac{\hCBB_{\l}-C_{\l}^{\rm L}}{C_{\l}^{\rm T}}
	\,,
}
where the quantities $\hCBB_{\l}$, $C_{\l}^{\rm L}$ and $C_{\l}^{\rm T}$ are 
the observed, lensing/delensed and tensor B-mode power spectrum, respectively. 
Here the spectrum $C_{\l}^{\rm T}$ is for $r=1$. 
Under the assumption that the covariance of the observed B-mode power spectrum is diagonal, 
the expected variance of the above quantity is given by
\al{
	\sigma^2_{\l} = \frac{2}{2\l+1}\left(\frac{C_{\l}^{\rm L}+rC_{\l}^{\rm T}}{C_{\l}^{\rm T}}\right)^2
	\,. 
}
This leads to an optimal estimator of $r$ as 
\al{
	\hr = (\sigma^r\rom{g})^2\sum_{\l}\frac{\hr_{\l}}{\sigma^2_{\l}}
	\,,
}
where $(\sigma^r\rom{g})^2$ is the expected variance of $\hr$: 
\al{
	(\sigma^r\rom{g})^2 \equiv \frac{1}{\sum_{\l}\sigma_{\l}^{-2}} = \left[\sum_{\l}\frac{2\l+1}{2}
		\left(\frac{C_{\l}^{\rm T}}{rC_{\l}^{\rm T}+C_{\l}^{\rm L}}\right)^2\right]^{-1}
	\,.
}
Note that, to see the impact of the non-Gaussian property of the B-modes on the estimation 
of $\hr$ clearly, we ignore the instrumental noise in the observed B-modes to be used for 
estimating $r$. 
The effect of the instrumental noise on the estimation of $\hr$ is discussed later. 
The estimated tensor-to-scalar ratio would be close to that obtained from the maximum likelihood method
\footnote{
Note that the estimator $\hr$ is not precisely the same as that obtained from the maximum likelihood 
method if the tensor-to-scalar ratio is close to zero and the fractional variance of $\hr$ is large. 
In our analysis here, the estimated values of the tensor-to-scalar ratio are much larger than 
the standard deviation. 
}
. 

The histograms of the estimated tensor-to-scalar ratio, which are obtained from $10000$ realizations 
of the tensor plus lensing and tensor plus delensed B-modes, are shown in Fig.~\ref{Fig:r1}. 
The input value of the tensor-to-scalar ratio is $0.05$. In both cases, the mean of 
the estimated tensor-to-scalar ratio equals to the input value within Monte Carlo error. 
Also in both cases, we find that the standard deviation of the tensor-to-scalar ratio increases 
only within a few percent ($\sim 2$\%) compared with that in the case of a Gaussian B-mode. 
Unlike the constraints on the power spectrum amplitude, 
the simulation results show that the standard deviation obtained in the presence of
the lensing-induced non-Gaussian B-modes ($\sigma^r_{\rm sim}$) 
is not significantly different from that in the case of a Gaussian B-mode ($\sigma^r_{\rm g}$). 
This is because well-defined estimators of $r$ usually extracts information mainly from 
large scale modes which earn a most part of the signal-to-noise of the tensor B-modes, 
while the B-modes on smaller scales, where non-Gaussian properties 
of the lensing/delensed B-modes are significant, are down-weighted in the estimators. 
Note that the value of $\sigma^r_{\rm sim}$ ($\sigma^r_{\rm g}$) is $0.00075$ ($0.00074$)
in the case of the lensing B-modes, and $0.00088$ ($0.00086$) in the case of the delensed B-modes, 
respectively. 
That is, $\sigma^r_{\rm sim}$ obtained in the case of the delensed B-modes is decreased 
by $\sim 15\%$ compared to that in the case of the lensing B-modes, 
while $\sigma^r_{\rm g}$ is decreased by $14\%$ after delensing. 
This means that the delensing efficiency is almost unchanged whether or not the non-Gaussian structure 
of the B-mode polarization is taken into account. 

Let us discuss the case if the instrumental noise is added in the lensing and delensed B-modes.
If a white noise is added, the B-modes at high multipoles behave like a Gaussian field, and 
the results of the estimation would be more close to those in the case of a Gaussian field. 
On the other hand, if the large scale B-modes are filtered out, the ratio of 
the standard deviations $\sigma^r_{\rm sim}/\sigma^r_{\rm g}$ would be increased. 
In practical situations, the large scale B-modes are significantly contaminated by 
Galactic foreground emission or 1/f noises. In Fig.~\ref{Fig:r2}, the ratio of 
the simulated standard deviation to that in the case of a Gaussian field 
($\sigma^r_{\rm sim}/\sigma^r\rom{g}$) is shown as a function of the minimum multipole included 
in the estimation ($\l\rom{min}$). As expected, the ratio increases for larger values of $\l\rom{min}$. 
Also, the figure compares the constraint on the tensor-to-scalar ratio in the case of 
the tensor plus lensing B-modes with that in the case of the tensor plus delensed B-modes. 
Note that, if $\l\rom{min}$ is increased to be larger than $100$, the discrepancy between 
the two cases grows further.

%% file: sec6.tex
\section{Summary and Discussion} \label{sec.6}

We have explored the non-Gaussian structure of the delensed B-mode polarization. 
Assuming an experiment of moderate sensitivity where the quadratic delensing is an almost optimal method, 
we find that not only the diagonal elements of the power spectrum covariance matrix but also its 
off-diagonal elements are reduced due to the delensing operation. 
In particular, the correlation coefficients between the power spectra at different multipoles 
become smaller than those of the original lensing B-modes. 
As a result, the constraint on the amplitude of the delensed B-mode power spectrum is not significantly degraded 
by the non-Gaussianity, compared to that on the amplitude of the lensing B-mode power spectrum. 
We show that the PDF of the power spectrum amplitude is broadened due to the non-Gaussianity, 
which leads to the increase of the variance, while the skewness and kurtosis of the PDF are basically consistent 
with those in the case of a Gaussian field. 
These features are well captured by the analytic model based on the 4th order Edgeworth expansion. 
Also, we show that the non-Gaussianity degrades the constraint on the tensor-to-scalar ratio after delensing 
within a few percent. 

In this paper, we assumed the noise level of $\Delta\rom{P}=6\mu$K-arcmin and the beam size of $\theta=4$ arcmin. 
In fact, our analytic model (and mathematical interpretation based on the model) 
works well as long as such moderate sensitivity experiments are assumed. 
For example, we checked that difference of $\sigma^A$ between simulations and the model prediction is 
$\sim 1$\% if $\Delta\rom{P}=9\mu$K-arcmin or $7.5\mu$K-arcmin, and $\sim 2$\% if $\Delta\rom{P}=3.5\mu$K-arcmin.
On the other hand, decreasing the noise level, 
we found that the analytic calculations based on the formulas described 
in Sec.~\ref{sec.4} eventually began to deviate from the simulation results. 
The ratio $\sigma^A/\sigma^A\rom{g}$ increases after initial decrease in the simulation, 
while that of the analytic model decreases monotonically. 
For example, if the noise level is $1.5\mu$K-arcmin, 
the analytic model overestimates the constraint on the power spectrum amplitude by $\sim 7$\%. 
In absence of instrumental noise (i.e., cosmic-variance limited up to $\l=2000$), 
the discrepancy reaches to $\sim 14$\%. 
This result implies that the approximations used in Sec.~\ref{sec.4} are no longer valid. 
In such high sensitivity experiments, since quadratic delensing is no longer an optimal method, 
iterative delensing proposed in Ref.~\cite{Hirata:2003ka} would be used for actual delensing analysis. 
Properties of the B-modes delensed in such analysis should be explored through another simulation 
(and another analytic modeling) based on the iterative method. 
It would be our future work.

%% file: appA.tex
\section{Statistical Properties of Power Spectrum Amplitude} 

\subsection{Mean} \label{sec:amp-mean}

Let us first derive the mean of the power spectrum amplitude $\hA$. 
The mean of the power spectrum amplitude is given by
\al{
	\ave{\hA} = \ave{\hA}\rom{g} + \ave{\hA k(\bm{a})}\rom{g}
	\,. 
}
Since $\ave{|a_{\l m}|^2}\rom{g}=C_{\l}$, the first term becomes 
\al{
	\ave{\hA}\rom{g} = \frac{\sum_{\l m}(\ave{|a_{\l m}|^2}\rom{g}/C_{\l})}{\sum_{\l}(2\l+1)} = 1
	\,. 
}
On the other hand, the second term vanishes. 
This is because, as described in Eq.~\eqref{Eq:deriv-k4}, $k(\bm{a})$ is a linear combination of 
four derivatives with respect to $a_{\l m}$. Therefore, for any $n$th order polynomial $f(a_{\l m})$ 
with $n<4$, a quantity $\ave{f(a_{\l m})k(\bm{a})}\rom{g}$ vanishes by integration by parts. 
The mean of $\hA$ then becomes $\ave{\hA}=1$.

\subsection{Variance} \label{sec:amp-var}

Next we derive the expression for the variance \eqref{Eq:sigmaA}, and also Eq.~\eqref{Eq:Cov-Full}.  
The variance of the amplitude estimator is defined as
\al{
	(\sigma^A)^2 \equiv \ave{\hA^2} - 1
		&= \frac{1}{[\sum_{\l}(2\l+1)]^2}\sum_{\l\l'}(2\l+1)(2\l'+1)(\ave{\hA_{\l}\hA_{\l'}}-1)
	\,.
}
Using the variance in the case of a Gaussian field $(\sigma^A\rom{g})^2$ defined in Eq.~\eqref{Eq:varA-g},
the above equation becomes Eq.~\eqref{Eq:sigmaA}:
\al{
	(\sigma^A)^2 &= \frac{(\sigma^A\rom{g})^2}{n_L}\sum_{\l\l'}
			\frac{(2\l+1)(2\l'+1)}{2}\frac{\ave{C_{\l}C_{\l'}} - C_{\l}C_{\l'}}{C_{\l}C_{\l'}} 
	\notag \\
		&= \frac{(\sigma^A\rom{g})^2}{n_L}\sum_{\l\l'}
			\frac{(2\l+1)(2\l'+1)}{2}\frac{{\rm Cov}_{\l\l'}}{C_{\l}C_{\l'}} 
	\,. 
}
Here, we denote the power spectrum covariance as 
\al{
	{\rm Cov}_{\l\l'} = \ave{C_{\l}C_{\l'}} - C_{\l}C_{\l'}
	\,. 
}
The power spectrum covariance is rewritten as
\al{
	{\rm Cov}_{\l\l'} = \ave{C_{\l}C_{\l'}}\rom{g} - C_{\l}C_{\l'} + \ave{C_{\l}C_{\l'}k(\bm{a})}\rom{g} 
	\,. 
}
The sum of the first two terms is the disconnected part of the power spectrum covariance:
\al{
	{\rm Cov}^{\rm g}_{\l\l'} \equiv \ave{C_{\l}C_{\l'}}\rom{g} - C_{\l}C_{\l'} 
		= \frac{2C_{\l}C_{\l'}}{2\l+1}\delta_{\l\l'}
	\,. \label{Eq:Covg}
}
On the other hand, the connected part of the covariance is related to the trispectrum as
(see equations from Eq.~\eqref{Eq:def-K} to Eq.~\eqref{Eq:Cov3} 
for the derivation of Eq.~\eqref{Eq:Cov2} from Eq.~\eqref{Eq:Cov1})
\al{
	{\rm Cov}_{\l\l'}^{\rm c} &\equiv \ave{C_{\l}C_{\l'}k(\bm{a})}\rom{g}
	\notag \\
		&= \frac{1}{(2\l+1)(2\l'+1)}\sum_{mm'} \ave{a_{\l m}a^*_{\l m}a_{\l' m'}a^*_{\l' m'}k(\bm{a})}\rom{g}
	\label{Eq:Cov1} \\
		&= \frac{1}{(2\l+1)(2\l'+1)}\sum_{mm'} (-1)^{m+m'}T^{\l\l\l'\l'}_{m,-m,m',-m'}
	\,. \label{Eq:Cov2}
}
The sum of the disconnected part \eqref{Eq:Covg} and connected part \eqref{Eq:Cov2} gives Eq.~\eqref{Eq:Cov-Full}. 

The remaining part of this subsection is devoted for derivation of Eq.~\eqref{Eq:Cov2} from Eq.~\eqref{Eq:Cov1}. 
Using Eq.~\eqref{Eq:deriv-k4}, we rewrite Eq.~\eqref{Eq:Cov1} as
\al{
	\frac{1}{(2\l+1)(2\l'+1)}\sum_{mm'} &\ave{a_{\l m}a^*_{\l m}a_{\l' m'}a^*_{\l' m'}k(\bm{a})}\rom{g}
	\notag \\ 
	&= \frac{1}{(2\l+1)(2\l'+1)}\sum_{mm'} \Int{}{\bm{a}}{_} a_{\l m}a^*_{\l m}a_{\l' m'}a^*_{\l' m'}k(\bm{a}) P\rom{g}(\bm{a})
	\notag \\ 
	&= \frac{1}{(2\l+1)(2\l'+1)}\sum_{mm'} 
		\frac{1}{24}\sum_{\l_i m_i}T^{\l_1\l_2\l_3\l_4}_{m_1m_2m_3m_4} K
	\,. \label{Eq:def-K}
}
where we define
\al{
	K \equiv \Int{}{\bm{a}}{_} a_{\l m}a^*_{\l m}a_{\l'm'}a^*_{\l'm'}
		\PD{}{a_{\l_1m_1}}\PD{}{a_{\l_2m_2}}\PD{}{a_{\l_3m_3}}\PD{}{a_{\l_4m_4}} P\rom{g}(\bm{a})
	\,. 
}
Introducing an operator $c_i$ which only affects on $a_{\l_i m_i}$ as $c_ia_{\l_i m_i}=a^*_{\l_i m_i}$, 
we perform integration by parts which gives 
\al{
	K &= -\Int{}{\bm{a}}{_} 
		[\delta_{\l\l_1}\delta_{mm_1} (a^*_{\l m}+c_1a_{\l m})a_{\l'm'}a^*_{\l'm'} + (\l m\leftrightarrow \l'm')]
	\notag \\
	&\qquad \times
		\PD{}{a_{\l_2m_2}}\PD{}{a_{\l_3m_3}}\PD{}{a_{\ell_4m_4}} P\rom{g}(\bm{a})
	\,.
}
Repeating integration by parts, we find
\al{
	K &= \Int{}{\bm{a}}{_} 
		[\delta_{\l\l_1}\delta_{mm_1}\delta_{\l\l_2}\delta_{mm_2}(c_2+c_1)a_{\l'm'}a^*_{\l'm'}
	\notag \\
		&\qquad + \delta_{\l\l_1}\delta_{mm_1} 
		(a^*_{\l m}+c_1a_{\l m})\delta_{\l'\l_2}\delta_{m'm_2}(a^*_{\l'm'}+c_2a_{\l'm'})
		+ (\l m\leftrightarrow \l'm')]
	\notag \\
	&\qquad \times
		\PD{}{a_{\l_3m_3}}\PD{}{a_{\l_4m_4}} P\rom{g}(\bm{a})
	\notag \\
	&= -\Int{}{\bm{a}}{_} 
		[\delta_{\l\l_1}\delta_{mm_1}\delta_{\l\l_2}\delta_{mm_2}(c_2+c_1)
			\delta_{\l'\l_3}\delta_{m'm_3}(a^*_{\l'm'}+c_3a_{\l'm'})
	\notag \\
		&\qquad + \delta_{\l\l_1}\delta_{mm_1}\delta_{\l\l_3}\delta_{mm_3}(c_3+c_1)
			\delta_{\l'\l_2}\delta_{m'm_2}(a^*_{\l'm'}+c_2a_{\l'm'})
	\notag \\
	&\qquad + \delta_{\l\l_1}\delta_{mm_1} (a^*_{\l m}+c_1a_{\l m})
			\delta_{\l'\l_2}\delta_{m'm_2}\delta_{\l'\l_3}\delta_{m'm_3}(c_3+c_2)]
		+ (\l m\leftrightarrow \l'm')]
	\notag \\
	&\qquad \times
		\PD{}{a_{\ell_4m_4}} P\rom{g}(\bm{a})
	\notag \\
	&= \Int{}{\bm{a}}{_} \frac{1}{4}\sum_{h,i,j,k=(1,2,3,4)}
			\delta_{\l\l_h}\delta_{mm_h}\delta_{\l\l_i}\delta_{mm_i}
			\delta_{\l'\l_j}\delta_{m'm_j}\delta_{\l'\l_k}\delta_{m'm_k}(c_h+c_i)(c_j+c_k) P\rom{g}(\bm{a})
	\notag \\
	&= \frac{1}{4}\sum_{h,i,j,k=(1,2,3,4)}
			\delta_{\l\l_h}\delta_{mm_h}\delta_{\l\l_i}\delta_{mm_i}
			\delta_{\l'\l_j}\delta_{m'm_j}\delta_{\l'\l_k}\delta_{m'm_k}(c_h+c_i)(c_j+c_k)
	\,.
}
The summation with respect to $h, i, j$ and $k$ is applied for every permutaion of $(1, 2, 3, 4)$.
Using the above equation, we obtain
\al{
	&\frac{1}{(2\l+1)(2\l'+1)}\sum_{mm'}
		\ave{a_{\l m}a^*_{\l m}a_{\l'm'}a^*_{\l'm'}k(\bm{a})}\rom{g}
	\notag \\
		&= \frac{1}{(2\l+1)(2\l'+1)}\sum_{mm'} (-1)^{m+m'}T^{\l\l\l'\l'}_{m,-m,m',-m'}
	\,. \label{Eq:Cov3}
}
This equation equals to Eq.~\eqref{Eq:Cov2}. 

\subsection{Probability distribution function of power spectrum amplitude} \label{sec:amp-PDF}

Here we derive the PDF of the power spectrum amplitude, $\hA$, shown in 
Eq.~\eqref{Eq:PDF-analytic}. The PDF of $\hA$ is defined as
\al{
	P(\hA) &\equiv \Int{n_L}{\bm{a}}{_} 
		\delta_{\rm D}\left(\hA-\frac{\sum_{\l m}|a_{\l m}|^2/C_{\l}}{n_L}\right)P(\bm{a})
	\notag \\
		 &\propto \Int{n_L}{\bm{a}}{_} 
		\delta_{\rm D}\left(\hA-\frac{\sum_{\l m}|a_{\l m}|^2/C_{\l}}{n_L}\right)
		[1+k(\bm{a})] \prod_{\l'} C^{-\frac{2\l'+1}{2}}_{\l'}
		\exp\left[-\sum_{m'=-\l'}^{\l'}\frac{|a_{\l' m'}|^2}{2C_{\l'}}\right]
	\notag \\
		& = \left[\prod_{\l'} C^{-\frac{2\l'+1}{2}}_{\l'}\right] \E^{-\frac{n_L}{2}\hA}
		\Int{n_L}{\bm{a}}{_} \delta_{\rm D}\left(\hA-\frac{\sum_{\l m}|a_{\l m}|^2/C_{\l}}{n_L}\right)
		[1+k(\bm{a})]
	\,. 
}
Here $\delta\rom{D}$ is the Dirac delta function.
Introducing variables $u_{\l m}$ so that $a_{\l m}=(n_L\hA C_{\l})^{1/2}u_{\l m}$, 
the above relation is simplified as
\al{
	P(\hA) &= P\rom{g}(\hA) 
		\Int{n_L}{\bm{u}}{_} \delta_{\rm D}\left(1-|\bm{u}|^2\right)[1+k(\bm{u},\hA)] 
	\notag \\ 
	&\equiv P\rom{g}(\hA) \mC{K}(\hA)
	\,, \label{Eq:appB:PDF}
}
where the PDF in the case of a Gaussian field has the dependence as
\al{
	P\rom{g}(\hA) \propto \hA^{\frac{n_L}{2}-1} \E^{-\frac{n_L}{2}\hA}
	\,. 
}
We define the correction factor as follows:
\al{
	\mC{K}(\hA) &\equiv \frac{P(\hA)}{P\rom{g}(\hA)} 
		= \Int{n_L}{\bm{u}}{_} \delta_{\rm D}\left(1-|\bm{u}|^2\right)
			+ \Int{n_L}{\bm{u}}{_} \delta_{\rm D}\left(1-|\bm{u}|^2\right)k(\bm{u},\hA)
	\,. \label{Eq:appB:K}
}

Let us consider a simplified expression for $\mC{K}(\hA)$. For this purpose, we use the following 
formula
\al{
	\Int{n_L}{\bm{x}}{_} \delta\rom{D} (1-|\bm{x}|^2) f(|\bm{x}|^2)
		&= S_{n_L} \Int{}{r}{_} r^{n_L-1} \delta\rom{D} (1-r^2) f(r^2)
	\notag \\
		&= \frac{S_{n_L}}{2}\Int{}{s}{_} s^{(n_L-2)/2} \delta\rom{D} (1-s) f(s) 
		= \frac{S_{n_L}f(1)}{2}
	\,,
}
where $S_{n_L}=2\pi^{n_L/2}/\Gamma(n_L/2)$ is the surface area of the unit sphere in $n_L-1$ dimension. 
From the above equation, the first term of Eq.~\eqref{Eq:appB:K} is given by $S_{n_L}/2$.
On the other hand, the evaluation of the second term requires a bit complicated calculation. 
Using the explicit expression for the kurtosis contribution \eqref{Eq:sum-k4}, 
we first rewrite the second term of Eq.~\eqref{Eq:appB:K} as
\al{
	\frac{1}{24}\sum_{\l_i m_i} 
		&\frac{T^{\l_1\l_2\l_3\l_4}_{m_1m_2m_3m_4}}{(C_{\l_1}C_{\l_2}C_{\l_3}C_{\l_4})^{1/2}}
	\Int{n_L}{\bm{u}}{_} \delta_{\rm D}\left(1-|\bm{u}|^2\right)
		\bigg(
			n_L^2\hA^2 u^*_{\l_1 m_1}u^*_{\l_2 m_2}u^*_{\l_3 m_3}u^*_{\l_4 m_4}
	\notag \\ 
		&\qquad - 6n_L\hA \delta_{\l_1\l_2}\delta_{-m_1,m_2}(-1)^{m_1}u^*_{\l_3 m_3}u^*_{\l_4 m_4}
	\notag \\
		&\qquad + 3\delta_{\l_1\l_2}\delta_{-m_1,m_2}\delta_{\l_3\l_4}\delta_{-m_3,m_4}(-1)^{m_1+m_3}
		\bigg)
	\,. \label{Eq:corr}
}
Note that this is a real number for a given set of $a_{\l m}$s. 
The above equation consists of the three terms: the term proportional to $\hA^2$, $\hA$ and 
independent of $\hA$. We first consider the term containing $\hA^2$. 
To simplify this term, 
we use the following general formula:
\al{
	&\sum_{ijkm}\mC{T}_{ijkm}\Int{n_L}{\bm{x}}{_} \delta_{\rm D}\left(1-|\bm{x}|^2\right) x_ix_jx_kx_m
	\notag \\
	&= \Int{n_L}{\bm{x}}{_} \delta_{\rm D}\left(1-|\bm{x}|^2\right) 
		\left(\sum_i \mC{T}_{iiii}|x_i|^4+3\sum_{i\not=j}\mC{T}_{ijij}|x_i|^2|x_j|^2\right)
	\notag \\
	&= 3I_4\sum_{ij}\mC{T}_{ijij}
	\,. \label{Eq:appB:4th}
}
Here, the quantity $\mC{T}_{ijkl}$ is assumed to be unchanged by the exchange of two idices, e.g. $\mC{T}_{ijkl}=\mC{T}_{jikl}$. 
We define $I_4$ as 
\al{
	I_4 \equiv \Int{n_L}{\bm{x}}{_} \delta_{\rm D}\left(1-|\bm{x}|^2\right) |x_i|^2|x_j|^2
	\,,
}
where $i \neq j$. Also, we use the fact that (see Sec.~\ref{sec:I4} for derivation):
\al{
	I_4 = \frac{1}{3}\Int{n_L}{\bm{x}}{_} \delta_{\rm D}\left(1-|\bm{x}|^2\right) |x_i|^4
	\,. \label{Eq:I4}
}
$I_4$ is independent of the indices ($i$ and $j$) appearing in the r.h.s. of the above equations. 
The quantity $I_4$ is computed as
\al{
	\left(3\sum_i+\sum_{i\not=j}\right)I_4
	&= \Int{n_L}{\bm{x}}{_} \delta_{\rm D}\left(1-|\bm{x}|^2\right) 
		\left(\sum_i|x_i|^4+\sum_{i\not=j}|x_i|^2|x_j|^2\right)
	\notag \\
	&= \Int{n_L}{\bm{x}}{_} \delta_{\rm D}\left(1-|\bm{x}|^2\right) \left(\sum_i|x_i|^2\right)^2
		= \frac{S_{n_L}}{2}
	\,. \label{Eq:I4}
}
By use of Eqs.~\eqref{Eq:appB:4th} and \eqref{Eq:I4}, the first term of Eq.~\eqref{Eq:corr} is given by
\al{
	& \frac{1}{24}\sum_{\l_i m_i}
		\frac{T^{\l_1\l_2\l_3\l_4}_{m_1m_2m_3m_4}}{(C_{\l_1}C_{\l_2}C_{\l_3}C_{\l_4})^{1/2}}
		\Int{n_L}{\bm{u}}{_} \delta_{\rm D}\left(1-|\bm{u}|^2\right)
		n_L^2\hA^2 u^*_{\l_1 m_1}u^*_{\l_2 m_2}u^*_{\l_3 m_3}u^*_{\l_4 m_4}
	\notag \\
	&= S_{n_L}\frac{n_L}{n_L+2}\frac{\hA^2}{16}\sum_{\l m\l'm'} 
		\frac{\ave{|a_{\l m}|^2|a_{\l' m'}|^2}_c}{C_{\l}C_{\l'}}
	\,. 
}
Next we consider the second term of Eq.~\eqref{Eq:corr}. Similarly as in the case of the first term, 
we define the following integral:
\al{
	I_2 \equiv \Int{n_L}{\bm{u}}{_} \delta_{\rm D}\left(1-|\bm{u}|^2\right) u_i u_j
		= \delta_{ij}\Int{n_L}{\bm{u}}{_} \delta_{\rm D}\left(1-|\bm{u}|^2\right) |u_i|^2
	\,.
}
This quantity satisfies
\al{
	\sum_i I_2 = \Int{n_L}{\bm{x}}{_} \delta_{\rm D}\left(1-|\bm{x}|^2\right) \sum_i|x_i|^2 = \frac{S_{n_L}}{2}
	\,. 
}
This reduces the second term of Eq.~\eqref{Eq:corr} to
\al{
	&\frac{-\hA}{4}\sum_{\l_i m_i} 
		T^{\l_1\l_2\l_3\l_4}_{m_1m_2m_3m_4}
		\Int{n_L}{\bm{u}}{_} \delta_{\rm D}(1-|\bm{u}|^2)
		\delta_{\l_1\l_2}\delta_{-m_1,m_2}(-1)^{m_1}u^*_{\l_3 m_3}u^*_{\l_4 m_4}
	\notag \\
	&= -S_{n_L}\frac{\hA}{8}\sum_{\l m\l'm'} \frac{\ave{|a_{\l m}|^2|a_{\l'm'}|^2}_c}{C_{\l}C_{\l'}}
	\,. 
}
Finally, the last term becomes
\al{
	&\frac{1}{8}\sum_{\l_i m_i} T^{\l_1\l_2\l_3\l_4}_{m_1m_2m_3m_4}
		\Int{n_L}{\bm{u}}{_} \delta_{\rm D}\left(1-|\bm{u}|^2\right)
		\delta_{\l_1\l_2}\delta_{-m_1,m_2}\delta_{\l_3\l_4}\delta_{-m_3,m_4}(-1)^{m_1+m_3}
	\notag \\
	&= \frac{S_{n_L}}{16}\sum_{\l m\l'm'} \frac{\ave{|a_{\l m}|^2|a_{\l'm'}|^2}_c}{C_{\l}C_{\l'}}
	\,. 
}
Combining these results, we find
\al{
	\mC{K}(\hA) = \frac{S_{n_L}}{2} + S_{n_L}\frac{(n_L/(n_L+2)\hA^2-2\hA+1)}{16}
		\sum_{\l m\l'm'} \frac{\ave{|a_{\l m}|^2|a_{\l'm'}|^2}_c}{C_{\l}C_{\l'}}
	\,.
}
We finally obtain the PDF of $\hA$ as 
\al{
	P(\hA) \propto \hA^{\frac{n_L}{2}-1} \E^{-\frac{n_L}{2}\hA} 
		\left(1+\frac{(n_L)/(n_L+2)\hA^2-2\hA+1}{8}
		\sum_{\l\l'}(2\l+1)(2\l'+1)\frac{{\rm Cov}^{\rm c}_{\l\l'}}{C_{\l}C_{\l'}}\right)
	\,. 
}
Note that the normalization is given by $(n_L/2)^{n_L/2}/\Gamma(n_L/2)$.

\subsection{Formula of integration on unit hypersurface} \label{sec:I4}

Finally, we derive the relationship between the following two integrals: 
\al{
	J_{ij} &\equiv \Int{n}{\bm{x}}{_} \delta_{\rm D}\left(1-|\bm{x}|^2\right) |x_i|^2|x_j|^2
	\,, \\
	J_i &\equiv \Int{n}{\bm{x}}{_} \delta_{\rm D}\left(1-|\bm{x}|^2\right) |x_i|^4
	\,.
}
In our derivation, we use
\al{
	F(\bm{t}) \equiv \Int{n}{\bm{x}}{_} \delta_{\rm D}\left(1-|\bm{x}|^2\right)
		\E^{-(t_1x_1^2+t_2x_2^2+\cdots+t_nx_n^2)}
	\,, \label{Eq:gfunc}
}
where $\bm{t}^t=(t_1,t_2,\cdots,t_n)$ are non-negative parameters.
This function is considered as a generating function of $J_{ij}$ and $J_i$, 
since these quantities are given as a derivative of $F(\bm{t})$ in terms of $t_i$: 
\al{
	J_{ij} &= \lim_{\bm{t}\to +0}\PD{^2 F(\bm{t})}{t_i\pd t_j}
	\,, &
	J_{i} &= \lim_{\bm{t}\to +0}\PD{^2 F(\bm{t})}{t_i^2}
	\,. \label{Eq:J}
}
The generating function \eqref{Eq:gfunc} is rewritten as
\al{
	F(\bm{t}) 
		&= \Int{n}{\bm{x}}{_} \Int{}{s}{_} \frac{1}{2\pi} \E^{-\iu s(1-|\bm{x}|^2)}
			\E^{-(t_1x_1^2+t_2x_2^2+\cdots+t_nx_n^2)}
	\notag \\ 
		&= \Int{}{s}{_} \frac{1}{2\pi} \E^{-\iu s} \prod_{i=1}^n \Int{}{x_i}{_} \exp\left[-(t_i-\iu s)x_i^2\right]
	\notag \\ 
		&= \Int{}{s}{_} \frac{1}{2\pi} \E^{-\iu s} \prod_{i=1}^n \left(\frac{\pi}{t_i-\iu s}\right)^{1/2}
	\,.
}
Substituting the above equation into Eq.~\eqref{Eq:J}, we find
\al{
	3J_{ij} = J_i
	\,.
}
Since $I_4=J_{ij}$, we obtain Eq.~\eqref{Eq:I4}.

%% file: appB.tex
\section{Delensed B-mode Power Spectrum Covariance} 

\subsection{Analytic expression} \label{sec:covBB}

Let us discuss analytic expression for the power spectrum covariance of the delensed B-mode polarization. 
The estimator of the delensed B-mode polarization is given by Eq.~\eqref{Eq:Quad-Delens}. 
For a realization of our simulation, the lensing potential is obtained from the simulated CMB map according to 
Eqs.~\eqref{Eq:estg-XY} and \eqref{Eq:estg}. 
The estimated lensing potential is decomposed into the lensing potential 
and the other remaining term (the noise of the lensing reconstruction) as 
\al{
	\estg_{\l m} = \grad_{\l m} + n_{\l m}
	\,. 
}
From Eq.~\eqref{Eq:Quad-LensB}, the estimator of the lensing B-mode is described as 
\al{
	\mS{B}_{\l m}[E,\grad^{\rm w}] + \mS{B}_{\l m}[E,n^{\rm w}]
	\,, 
}
where the Wiener filtered multipoles $\grad^{\rm w}_{\l m}$ and $n^{\rm w}_{\l m}$ are defined 
by multiplying the Wiener filter $W^{\grad}_{\l}$ in Eq.~\eqref{Eq:WF-grad} 
to the $\grad_{\l m}$ and $n_{\l m}$, respectively. 
The dominant contribution of the lensing B-mode at large scales is also expressed by the convolution 
of the E-mode polarization and lensing potential as described in Eq.~\eqref{Eq:Lensing-E-to-B} \cite{Smith:2010gu}. 
On the assumption that the lensing B-mode is given by Eq.~\eqref{Eq:Lensing-E-to-B}, 
the delensed B-mode polarization (in the absence of the tensor perturbations) is described 
as a sum of two components: 
\al{
	\rB_{\l m} = \mS{B}_{\l m}[E,(1-W^{\grad})\grad] - \mS{B}_{\l m}[E,n^{\rm w}]
	\,. \label{Eq:Delensed-B-lin}
}
In the above equations, we assume that the Wiener filter for the E-mode ($W^{\rm E}$) is unity up to $\l=2000$, 
i.e., uncertainty in the E-mode polarization is dominated by its cosmic variance and the instrumental noise is 
negligible. This approximation would be valid for ongoing and future high-resolution experiments 
since their polarization sensitivity will be better than $\mC{O}(1)\mu$K-arcmin also 
with small beam sizes of a few arcminute. 
The filter function for the lensing potential ($W^{\grad}_\l$) depends on the lensing power spectrum of 
the theoretical model which is identified through a parameter estimation procedure in actual analysis. 
We assume that the model is correctly identified in our simulation and 
the filter function is henceforth treated as a quantity independent of $\Cgg_\l$. 

The delensed B-mode power spectrum is computed from Eq.~\eqref{Eq:Delensed-B-lin}. 
Under the assumption that the lensing potential $\grad_{\l m}$ is statistically independent from 
the residual reconstruction noise $n_{\l m}$, 
the delensed B-mode power spectrum becomes \cite{Smith:2010gu}
\al{
	\rCBB_{\l} 
		&\simeq \Xi_{\l} [\CEE,(1-W^{\grad})^2\Cgg] + \Xi_{\l}[\CEE,(W^{\grad})^2\Ngg]
	\label{Eq:rCBB-lin} \\
		&\simeq \Xi_{\l} [\CEE,(1-W^{\grad})\Cgg] 
	\,, \label{Eq:rCBB-reduced}
}
where $\Ngg_\l$ is the power spectrum of $n_{\l m}$, and, from Eq.~\eqref{Eq:rCBB-lin} to 
\eqref{Eq:rCBB-reduced}, we assume that $\Ngg_{\l}$ corresponds to $A^{\grad}_{\l}$ given 
in Eq.~\eqref{Eq:AL_GRAD}. 
Note that, strictly speaking, the lensing potential and reconstruction noise are correlated 
because the reconstruction noise has contributions from the lensing potential and lensed E-mode. 
The impact of the correlation between $\grad_{\l m}$ and $n_{\l m}$ would be, however, negligible 
at least for experiments assumed in this paper, since 
the approximate form of Eq.~\eqref{Eq:rCBB-reduced} is in good agreement with the simulation results. 
This issue is also discussed in our previous work \cite{Namikawa:2014yca} 
where we assumed the similar experimental specification. 

To derive the power spectrum covariance in the delensed case, 
we need to compute the correlation of Eq.~\eqref{Eq:rCBB-lin}. 
Denoting the first and second terms of Eq.~\eqref{Eq:rCBB-lin} as
\al{
	D^{\grad}_{\l} &\equiv \Xi_{\l} [\CEE,(1-W^{\grad})^2\Cgg] 
	\,, \\ 
	D^{n}_{\l} &\equiv \Xi_{\l}[\CEE,(W^{\grad})^2\Ngg] 
	\,, 
}
we write the covariance of the delensed B-mode power spectrum as the sum of the following three terms: 
\al{
	T^{4\grad}_{\l\l'} &= \ave{D^{\grad}_{\l}D^{\grad}_{\l'}} 
	\,, \\ 
	T^{4n}_{\l\l'} &= \ave{D^n_{\l}D^n_{\l'}} 
	\,, \\
	T^{2\grad 2n}_{\l\l'} &= \ave{D^{\grad}_{\l}D^n_{\l'}} + \text{($\l\leftrightarrow\l'$)}
	\,. 
}
The first two terms would be evaluated analytically with the analogy used in 
the case of the lensing B-modes \cite{Li:2006pu,2012:Benoit}, i.e., 
\al{
	T^{4\grad}_{\l\l'}
	&= \sum_L \frac{2}{2L+1} 
		\left\{ \PD{D^{\grad}_{\l}}{\ln\CEE_L}\PD{D^{\grad}_{\l'}}{\ln\CEE_L}
		+ \PD{D^{\grad}_{\l}}{\ln \Cgg_L}\PD{D^{\grad}_{\l'}}{\ln \Cgg_L}
		\right\}
	\notag \\ 
	&= \sum_L \frac{2}{2L+1} 
		\left\{ \PD{D^{\grad}_{\l}}{\ln\CEE_L}\PD{D^{\grad}_{\l'}}{\ln\CEE_L}
		+ (1-W^{\grad}_L)^4\PD{\tCBB_{\l}}{\ln\Cgg_L}\PD{\tCBB_{\l'}}{\ln\Cgg_L}
		\right\}
	\,, 
}
and 
\al{
	T^{4n}_{\l\l'} = \sum_L \frac{2}{2L+1} 
		\left\{ \PD{D^n_{\l}}{\ln\CEE_L}\PD{D^n_{\l'}}{\ln\CEE_L}
		+ \left[\frac{(W^{\grad}_L)^2\Ngg_L}{\Cgg_L}\right]^2
			\PD{\tCBB_{\l}}{\ln\Cgg_L}\PD{\tCBB_{\l'}}{\ln\Cgg_L}
		\right\}
	\,. 
}
Here we assume that the power spectra $\CEE$, $\Cgg$ and $\Ngg$ have Gaussian covariance. 
Note that, as discussed in Refs.~\cite{Schmittfull:2013uea,Ade:2013tyw,Ade:2015zua}, 
the covariance between $\CEE$, $\Cgg$ and $\Ngg$ can have off-diagonal elements and is shown that 
these are basically negligible. The remaining term $T^{2\grad 2n}_{\l\l'}$ would be approximated as 
\al{
	T^{2\grad 2n}_{\l\l'} 
		&= \sum_L \frac{2}{2L+1} 
		\bigg\{ \PD{D^{\grad}_{\l}}{\ln\CEE_L}\PD{D^n_{\l'}}{\ln\CEE_L}
	\notag \\
		&\qquad + (1-W^{\grad}_L)^2(W^{\grad}_L)^2 \frac{\Ngg_L}{\Cgg_L}
		\PD{\tCBB_{\l}}{\ln \Cgg_L}\PD{\tCBB_{\l'}}{\ln \Cgg_L}
		\bigg\}
		+ \text{($\l\leftrightarrow\l'$)}
	\,. 
}
By combining the above three terms, we find 
\al{
	T^{4\grad}_{\l\l'} + T^{4n}_{\l\l'} + T^{2\grad 2n}_{\l\l'}
		&= \sum_L \frac{2}{2L+1} 
		\bigg\{ \PD{(D^{\grad}_{\l}+D^n_{\l})}{\ln\CEE_L}\PD{(D^{\grad}_{\l'}+D^n_{\l'})}{\ln\CEE_L}
	\notag \\
		&\qquad + \left[(1-W^{\grad}_L)^2+\frac{(W^{\grad}_L)^2\Ngg_L}{\Cgg_L}\right]^2
			\PD{\tCBB_{\l}}{\ln\Cgg_L}\PD{\tCBB_{\l'}}{\ln\Cgg_L}
		\bigg\}
	\notag \\
		&\simeq \sum_L \frac{2}{2L+1} 
		\bigg\{ \PD{(D^{\grad}_{\l}+D^n_{\l})}{\ln\CEE_L}\PD{(D^{\grad}_{\l'}+D^n_{\l'})}{\ln\CEE_L}
	\notag \\
		&\qquad + (1-W_L^{\grad})^2
			\PD{\tCBB_{\l}}{\ln\Cgg_L}\PD{\tCBB_{\l'}}{\ln\Cgg_L}
		\bigg\}
	\notag \\
		&\simeq \sum_L \frac{2}{2L+1} 
		\bigg\{ \PD{\rCBB_{\l}}{\ln\CEE_L}\PD{\rCBB_{\l'}}{\ln\CEE_L}
		+ \PD{\rCBB_{\l}}{\ln\Cgg_L}\PD{\rCBB_{\l'}}{\ln\Cgg_L}
		\bigg\}
	\,.
}
From the first to second equation, we assume that $\Ngg_\l$ corresponds to $A^{\grad}$, and 
from the second to third equation, we use Eq.~\eqref{Eq:rCBB-reduced}.
The covariance of the delensed B-mode power spectrum given above equals to the off-diagonal part of 
Eq.~\eqref{Eq:CovBB-d}. Note that we compute the derivative of the delensed B-mode power spectrum 
$\rCBB_{\l}$ with respect to $\Cgg_\l$ via Eq.~\eqref{Eq:rCBB-reduced}. 

\subsection{Numerical computation} \label{Sec:Comp-Cov}

Here we describe our method of computing the power spectrum covariance of the lensing and delensed B-mode. 
The power spectrum covariance is given by
\al{
	{\rm Cov}_{\l\l'}^{\rm BB} &= \frac{2}{2\l+1}C_{\l}^2\delta_{\l\l'}
		+ {\rm Cov}^{\rm E}_{\l\l'} + {\rm Cov}^{\grad}_{\l\l'}
	\,,
}
where we denote the connected part of the covariance as 
\al{
	{\rm Cov}^{\rm E}_{\l\l'} &=
		\sum_L \PD{C_{\l}}{\CEE_L}\frac{2(\CEE_L)^2}{2L+1}\PD{C_{\l'}}{\CEE_L}
	\,, \\
	{\rm Cov}^{\grad}_{\l\l'} &= 
		\sum_L \PD{C_{\l}}{\Cgg_L}\frac{2(\Cgg_L)^2}{2L+1}\PD{C_{\l'}}{\Cgg_L}
	\,. 
}
To evaluate the connected part of the covariance, Cov$^{\rm E}$ and Cov$^{\grad}$, 
we rewrite the derivatives as 
\al{
	\PD{C_{\l}}{\CEE_L} &= \Xi^{\grad}_{\l L}[\Cgg]
	\,, \\ 
	\PD{C_{\l}}{\Cgg_L} &= \Xi^{\rm E}_{\l L}[\CEE]
	\,.
}
Here we define
\al{
	\Xi^{\grad}_{\l L}[A] \equiv \frac{1}{2\l+1}\sum_{L'} (\mC{S}^{(-)}_{\l LL'})^2 A_{L'} 
	\,, \\ 
	\Xi^{\rm E}_{\l L}[A] \equiv \frac{1}{2\l+1}\sum_{L'} (\mC{S}^{(-)}_{\l L'L})^2 A_{L'} 
	\,. 
}
Note that
\al{
	\Xi_{\l}[A,B] = \sum_L A_L\Xi^{\grad}_{\l L}[B] = \sum_L B_L\Xi^{\rm E}_{\l L}[A] 
	\,. 
}
For instance, ${\rm Cov}^{\rm E}_{\l\l'}$ is then given by 
\al{
	{\rm Cov}^{\rm E}_{\l\l'}
		&= \sum_L \PD{C_{\l}}{\CEE_L}\frac{2(\CEE_L)^2}{2L+1}\PD{C_{\l'}}{\CEE_L}
	\notag \\
		&= \sum_L \Xi^{\grad}_{\l L}[\Cgg] \frac{2(\CEE_L)^2}{2L+1} \Xi^{\grad}_{\l'L}[\Cgg]
	\,.
}
Denoting 
\al{
	F_L^{\l'} = \frac{2(\CEE_L)^2}{2L+1}\Xi^{\grad}_{\l'L}[\Cgg]
	\,,
}
we obtain
\al{
	{\rm Cov}^{\rm E}_{\l\l'} = \Xi_{\l}[F^{\l'},\Cgg]
	\,.
}
Similarly, we find 
\al{
	{\rm Cov}^{\grad}_{\l\l'} = \Xi_{\l}[\CEE,G^{\l'}]
	\,,
}
where we define
\al{
	G_L^{\l'} = \frac{2(\Cgg_L)^2}{2L+1}\Xi^{\rm E}_{\l'L}[\CEE]
	\,.
}
The summations in $\Xi_{\l}$, $\Xi_{\l L}^{\rm E}$ and $\Xi_{\l L}^{\grad}$ are efficiently evaluated by use of 
the reduced wigner d functions as described in Ref.~\cite{Smith:2010gu}.
The power spectrum covariance of the delensed B-modes is also evaluated in the above similar manner.

%% file: ms.bbl
\providecommand{\href}[2]{#2}\begingroup\raggedright\begin{thebibliography}{10}

\bibitem{Ade:2015tva}
{\bf BICEP2 and PLANCK Collaboration} , {\it ``A Joint Analysis of BICEP2/Keck
  Array and Planck Data''},  {\em Phys. Rev. Lett.} (2015)
  [\href{http://arxiv.org/abs/1502.00612}{{\tt arXiv:1502.00612}}].

\bibitem{Ade:2015xua}
{\bf Planck Collaboration} , {\it ``Planck 2015 results. XIII. Cosmological
  parameters''},  \href{http://arxiv.org/abs/1502.01589}{{\tt
  arXiv:1502.01589}}.

\bibitem{Dunkley:2008am}
J.~Dunkley {\em et~al.}, {\it ``CMBPol Mission Concept Study: Prospects for
  polarized foreground removal''},  {\em AIP Conf. Proc.} {\bf 1141} (2009)
  222, [\href{http://arxiv.org/abs/0811.3915}{{\tt arXiv:0811.3915}}].

\bibitem{Betoule:2009}
M.~Betoule, E.~Pierpaoli, J.~Delabrouille, M.~Le~Jeune, and J.-F. Cardoso, {\it
  ``Measuring the tensor to scalar ratio from CMB B-modes in the presence of
  foregrounds''},  {\em Astronomy and Astrophysics} {\bf 503} (sep, 2009)
  691--706, [\href{http://arxiv.org/abs/0901.1056}{{\tt arXiv:0901.1056}}].

\bibitem{Ichiki:2014}
K.~Ichiki, {\it ``CMB foreground: A concise review''},  {\em Prog. Theor. Exp.
  Phys.} {\bf 06} (2014) B109.

\bibitem{Katayama:2011eh}
N.~Katayama and E.~Komatsu, {\it ``Simple foreground cleaning algorithm for
  detecting primordial B-mode polarization of the cosmic microwave
  background''},  {\em Astrophys. J.} {\bf 737} (2011) 78,
  [\href{http://arxiv.org/abs/1101.5210}{{\tt arXiv:1101.5210}}].

\bibitem{Zaldarriaga:1998ar}
M.~Zaldarriaga and U.~Seljak, {\it ``Gravitational lensing effect on cosmic
  microwave background polarization''},  {\em Phys. Rev. D} {\bf 58} (1998)
  023003, [\href{http://arxiv.org/abs/astro-ph/9803150}{{\tt
  astro-ph/9803150}}].

\bibitem{Lewis:2006fu}
A.~Lewis and A.~Challinor, {\it ``Weak gravitational lensing of the CMB''},
  {\em Phys. Rept.} {\bf 429} (2006) 1--65,
  [\href{http://arxiv.org/abs/astro-ph/0601594}{{\tt astro-ph/0601594}}].

\bibitem{Seljak:2003pn}
U.~Seljak and C.~M. Hirata, {\it ``Gravitational lensing as a contaminant of
  the gravity wave signal in CMB''},  {\em Phys. Rev. D} {\bf 69} (2004)
  043005, [\href{http://arxiv.org/abs/astro-ph/0310163}{{\tt
  astro-ph/0310163}}].

\bibitem{Smith:2010gu}
K.~M. Smith {\em et~al.}, {\it ``Delensing CMB Polarization with External
  Datasets''},  {\em JCAP} {\bf 1206} (2012) 014,
  [\href{http://arxiv.org/abs/1010.0048}{{\tt arXiv:1010.0048}}].

\bibitem{Boyle:2014kba}
L.~Boyle, K.~M. Smith, C.~Dvorkin, and N.~Turok, {\it ``On testing and
  extending the inflationary consistency relation for tensor modes''},
  \href{http://arxiv.org/abs/1408.3129}{{\tt arXiv:1408.3129}}.

\bibitem{Simard:2014aqa}
G.~Simard, D.~Hanson, and G.~Holder, {\it ``Prospects for Delensing the Cosmic
  Microwave Background for Studying Inflation''},
  \href{http://arxiv.org/abs/1410.0691}{{\tt arXiv:1410.0691}}.

\bibitem{Namikawa:2014lla}
T.~Namikawa, D.~Yamauchi, and A.~Taruya, {\it ``Future detectability of
  gravitational-wave induced lensing from high-sensitivity CMB experiments''},
  {\em Phys. Rev. D} {\bf 91} (2015), no.~4 043531,
  [\href{http://arxiv.org/abs/1411.7427}{{\tt arXiv:1411.7427}}].

\bibitem{Teng:2011xc}
W.-H. Teng, C.-L. Kuo, and J.-H.~P. Wu, {\it ``Cosmic Microwave Background
  Delensing Revisited: Residual Biases and a Simple Fix''},
  \href{http://arxiv.org/abs/1102.5729}{{\tt arXiv:1102.5729}}.

\bibitem{Kamada:2014qta}
K.~Kamada, Y.~Miyamoto, D.~Yamauchi, and J.~Yokoyama, {\it ``Effects of cosmic
  strings with delayed scaling on CMB anisotropy''},  {\em Phys. Rev. D} {\bf
  90} (2014), no.~8 083502, [\href{http://arxiv.org/abs/1407.2951}{{\tt
  arXiv:1407.2951}}].

\bibitem{Saltas:2014dha}
I.~D. Saltas, I.~Sawicki, L.~Amendola, and M.~Kunz, {\it ``Anisotropic stress
  as signature of non-standard propagation of gravitational waves''},
  \href{http://arxiv.org/abs/1406.7139}{{\tt arXiv:1406.7139}}.

\bibitem{Fenu:2009JCAP}
E.~Fenu, D.~G. Figueroa, R.~Durrer, and J.~Garcia-Bellido, {\it ``Gravitational
  waves from self-ordering scalar fields''},  {\em JCAP} {\bf 10} (oct, 2009)
  5, [\href{http://arxiv.org/abs/0908.0425}{{\tt arXiv:0908.0425}}].

\bibitem{Figueroa:2013PRL}
D.~G. Figueroa, M.~Hindmarsh, and J.~Urrestilla, {\it ``Exact Scale-Invariant
  Background of Gravitational Waves from Cosmic Defects''},  {\em Physical
  Review Letters} {\bf 110} (mar, 2013) 101302,
  [\href{http://arxiv.org/abs/1212.5458}{{\tt arXiv:1212.5458}}].

\bibitem{vanEngelen:2014zlh}
{\bf ACT Collaboration} , A.~van Engelen {\em et~al.}, {\it ``The Atacama
  Cosmology Telescope: Lensing of CMB Temperature and Polarization Derived from
  Cosmic Infrared Background Cross-Correlation''},
  \href{http://arxiv.org/abs/1412.0626}{{\tt arXiv:1412.0626}}.

\bibitem{Ade:2015zua}
{\bf Planck Collaboration} , {\it ``Planck 2015 results. XV. Gravitational
  lensing''},  \href{http://arxiv.org/abs/1502.01591}{{\tt arXiv:1502.01591}}.

\bibitem{PB1:2013a}
{\bf POLARBEAR Collaboration} , {\it ``Measurement of the Cosmic Microwave
  Background Polarization Lensing Power Spectrum with the POLARBEAR
  experiment''},  {\em Phys.Rev.Lett.} {\bf 113} (2014) 021301,
  [\href{http://arxiv.org/abs/1312.6646}{{\tt arXiv:1312.6646}}].

\bibitem{Hanson:2013daa}
{\bf SPT Collaboration} , D.~Hanson {\em et~al.}, {\it ``Detection of B-mode
  Polarization in the Cosmic Microwave Background with Data from the South Pole
  Telescope''},  {\em Phys. Rev. Lett.} {\bf 111} (2013) 141301,
  [\href{http://arxiv.org/abs/1307.5830}{{\tt arXiv:1307.5830}}].

\bibitem{Story:2014hni}
{\bf SPT Collaboration} , K.~T. Story {\em et~al.}, {\it ``A Measurement of the
  Cosmic Microwave Background Gravitational Lensing Potential from 100 Square
  Degrees of SPTpol Data''},  \href{http://arxiv.org/abs/1412.4760}{{\tt
  arXiv:1412.4760}}.

\bibitem{Calabrese:2014gwa}
E.~Calabrese {\em et~al.}, {\it ``Precision Epoch of Reionization studies with
  next-generation CMB experiments''},  {\em JCAP} {\bf 1408} (2014) 010,
  [\href{http://arxiv.org/abs/1406.4794}{{\tt arXiv:1406.4794}}].

\bibitem{SimonsArray}
K.~Arnold {\em et~al.}, {\it ``The Simons Array: expanding POLARBEAR to three
  multi-chroic telescopes''},  {\em Proc. SPIE} {\bf 91531} (2014) 91531F.

\bibitem{Benson:2014}
B.~A. Benson, {\it ``SPT-3G: A Next-Generation Cosmic Microwave Background
  Polarization Experiment on the South Pole Telescope''},  {\em Proceedings of
  SPIE} (2014) [\href{http://arxiv.org/abs/1407.2973}{{\tt arXiv:1407.2973}}].

\bibitem{Abazajian:2013vfg}
K.~Abazajian {\em et~al.}, {\it ``Inflation Physics from the Cosmic Microwave
  Background and Large Scale Structure''},  {\em Astropart. Phys.} {\bf 63}
  (2015) 55--65, [\href{http://arxiv.org/abs/1309.5381}{{\tt
  arXiv:1309.5381}}].

\bibitem{Sherwin:2015baa}
B.~D. Sherwin and M.~Schmittfull, {\it ``Delensing the CMB with the Cosmic
  Infrared Background''},  \href{http://arxiv.org/abs/1502.05356}{{\tt
  arXiv:1502.05356}}.

\bibitem{Smith:2004up}
K.~M. Smith, W.~Hu, and M.~Kaplinghat, {\it ``Weak lensing of the CMB: Sampling
  errors on B-modes''},  {\em Phys. Rev. D} {\bf 70} (2004) 043002,
  [\href{http://arxiv.org/abs/astro-ph/0402442}{{\tt astro-ph/0402442}}].

\bibitem{Smith:2006nk}
K.~M. Smith, W.~Hu, and M.~Kaplinghat, {\it ``Cosmological Information from
  Lensed CMB Power Spectra''},  {\em Phys. Rev. D} {\bf 74} (2006) 123002,
  [\href{http://arxiv.org/abs/astro-ph/0607315}{{\tt astro-ph/0607315}}].

\bibitem{Li:2006pu}
C.~Li, T.~L. Smith, and A.~Cooray, {\it ``Non-Gaussian Covariance of CMB
  B-modes of Polarization and Parameter Degradation''},  {\em Phys. Rev. D}
  {\bf 75} (2007) 083501, [\href{http://arxiv.org/abs/astro-ph/0607494}{{\tt
  astro-ph/0607494}}].

\bibitem{2012:Benoit}
A.~Benoit-L\'evy, K.~M. Smith, and W.~Hu, {\it ``Non-Gaussian structure of the
  lensed CMB power spectra covariance matrix''},  {\em Phys. Rev. D} {\bf 86}
  (2012), no.~12 123008, [\href{http://arxiv.org/abs/1205.0474}{{\tt
  arXiv:1205.0474}}].

\bibitem{Ade:2013zuv}
{\bf Planck Collaboration} , {\it ``Planck 2013 results. XVI. Cosmological
  parameters''},  {\em Astron. Astrophys.} {\bf 571} (2014) A16,
  [\href{http://arxiv.org/abs/1303.5076}{{\tt arXiv:1303.5076}}].

\bibitem{Namikawa:2014yca}
T.~Namikawa and R.~Nagata, {\it ``Lensing reconstruction from a patchwork of
  polarization maps''},  {\em JCAP} {\bf 1409} (2014) 009,
  [\href{http://arxiv.org/abs/1405.6568}{{\tt arXiv:1405.6568}}].

\bibitem{Namikawa:2011cs}
T.~Namikawa, D.~Yamauchi, and A.~Taruya, {\it ``Full-sky lensing reconstruction
  of gradient and curl modes from CMB maps''},  {\em JCAP} {\bf 1201} (2012)
  007, [\href{http://arxiv.org/abs/1110.1718}{{\tt arXiv:1110.1718}}].

\bibitem{Lewis:1999bs}
A.~Lewis, A.~Challinor, and A.~Lasenby, {\it ``Efficient Computation of CMB
  anisotropies in closed FRW models''},  {\em Astrophys. J.} {\bf 538} (2000)
  473--476, [\href{http://arxiv.org/abs/astro-ph/9911177}{{\tt
  astro-ph/9911177}}].

\bibitem{Challinor:2005jy}
A.~Challinor and A.~Lewis, {\it ``Lensed CMB power spectra from all-sky
  correlation functions''},  {\em Phys. Rev. D} {\bf 71} (2005) 103010,
  [\href{http://arxiv.org/abs/astro-ph/0502425}{{\tt astro-ph/0502425}}].

\bibitem{Hu:2000ee}
W.~Hu, {\it ``Weak lensing of the CMB: A harmonic approach''},  {\em Phys. Rev.
  D} {\bf 62} (2000) 043007, [\href{http://arxiv.org/abs/astro-ph/0001303}{{\tt
  astro-ph/0001303}}].

\bibitem{Okamoto:2003zw}
T.~Okamoto and W.~Hu, {\it ``CMB Lensing Reconstruction on the Full Sky''},
  {\em Phys. Rev. D} {\bf 67} (2003) 083002,
  [\href{http://arxiv.org/abs/astro-ph/0301031}{{\tt astro-ph/0301031}}].

\bibitem{Hanson:2010rp}
D.~Hanson {\em et~al.}, {\it ``CMB temperature lensing power reconstruction''},
   {\em Phys. Rev. D} {\bf 83} (2011) 043005,
  [\href{http://arxiv.org/abs/1008.4403}{{\tt arXiv:1008.4403}}].

\bibitem{Lewis:2011fk}
A.~Lewis, A.~Challinor, and D.~Hanson, {\it ``The shape of the CMB lensing
  bispectrum''},  {\em JCAP} {\bf 1103} (2011) 018,
  [\href{http://arxiv.org/abs/1101.2234}{{\tt arXiv:1101.2234}}].

\bibitem{Hirata:2003ka}
C.~M. Hirata and U.~Seljak, {\it ``Reconstruction of lensing from the cosmic
  microwave background polarization''},  {\em Phys. Rev. D} {\bf 68} (2003)
  083002, [\href{http://arxiv.org/abs/astro-ph/0306354}{{\tt
  astro-ph/0306354}}].

\bibitem{Hamimeche:2008ai}
S.~Hamimeche and A.~Lewis, {\it ``Likelihood Analysis of CMB Temperature and
  Polarization Power Spectra''},  {\em Phys. Rev. D} {\bf 77} (2008) 103013,
  [\href{http://arxiv.org/abs/0801.0554}{{\tt arXiv:0801.0554}}].

\bibitem{2010:Regan}
D.~M. Regan, E.~P.~S. Shellard, and J.~R. Fergusson, {\it ``General CMB and
  primordial trispectrum estimation''},  {\em Phys. Rev. D} {\bf 82} (jul,
  2010) 023520, [\href{http://arxiv.org/abs/1004.2915}{{\tt arXiv:1004.2915}}].

\bibitem{Namikawa:2012pe}
T.~Namikawa, D.~Hanson, and R.~Takahashi, {\it ``Bias-Hardened CMB Lensing''},
  {\em Mon. Not. Roy. Astron. Soc.} {\bf 431} (2013) 609--620,
  [\href{http://arxiv.org/abs/1209.0091}{{\tt arXiv:1209.0091}}].

\bibitem{Namikawa:2013xka}
T.~Namikawa and R.~Takahashi, {\it ``Bias-Hardened CMB Lensing with
  Polarization''},  {\em Mon. Not. Roy. Astron. Soc.} {\bf 438} (2014), no.~2
  1507--1517, [\href{http://arxiv.org/abs/1310.2372}{{\tt arXiv:1310.2372}}].

\bibitem{Gorski:2004by}
K.~Gorski {\em et~al.}, {\it ``HEALPix - A Framework for high resolution
  discretization, and fast analysis of data distributed on the sphere''},  {\em
  Astrophys. J.} {\bf 622} (2005) 759--771,
  [\href{http://arxiv.org/abs/astro-ph/0409513}{{\tt astro-ph/0409513}}].

\bibitem{Schmittfull:2013uea}
M.~M. Schmittfull, A.~Challinor, D.~Hanson, and A.~Lewis, {\it ``Joint analysis
  of CMB temperature and lensing-reconstruction power spectra''},  {\em Phys.
  Rev. D} {\bf 88} (2013), no.~6 063012,
  [\href{http://arxiv.org/abs/1308.0286}{{\tt arXiv:1308.0286}}].

\bibitem{Ade:2013tyw}
{\bf Planck Collaboration} , {\it ``Planck 2013 results. XVII. Gravitational
  lensing by large-scale structure''},  {\em Astron. Astrophys.} {\bf 571}
  (2014) A17, [\href{http://arxiv.org/abs/1303.5077}{{\tt arXiv:1303.5077}}].

\end{thebibliography}\endgroup
